\newcommand{\bx}{{\bf x}}
\newcommand{\bl}{\widehat{\bf s}}
\newcommand{\fts}{{\widetilde{s}}}
\newcommand{\ftr}{{\widetilde{r}}}
\newcommand{\ftv}{{\widetilde{v}}}

\newcommand{\bb}{{\bf b}}
\newcommand{\ftR}{{\widetilde{R}}}
\newcommand{\Aeff}{A_{\text{eff}}}

\documentclass[twocolumn]{emulateapj}
\usepackage[breaklinks,colorlinks,urlcolor=blue,citecolor=blue,linkcolor=blue]{hyperref}

\usepackage{amsmath}
\usepackage{graphicx}
\usepackage{natbib}
\usepackage{color}
\usepackage{hyperref}
\usepackage{comment}
\begin{document}
\shorttitle{The HERA Dish II: Electromagnetic Simulations and Science Implications}
\shortauthors{Ewall-Wice et al.}

\title{The Hydrogen Epoch of Reionization Array Dish II: Characterization of Spectral Structure with Electromagnetic  Simulations and its science Implications.}
\author{
Aaron Ewall-Wice \altaffilmark{1,2},
Richard Bradley\altaffilmark{3,4},
David Deboer\altaffilmark{5},
Jacqueline Hewitt\altaffilmark{1,2},
Aaron Parsons\altaffilmark{5},
James Aguirre\altaffilmark{6},
Zaki S. Ali\altaffilmark{5},
Judd Bowman\altaffilmark{7},
Carina Cheng\altaffilmark{5},
Abraham R. Neben\altaffilmark{1,2},
Nipanjana Patra\altaffilmark{5},
Nithyanandan Thyagarajan \altaffilmark{7},
Mariet Venter\altaffilmark{8,9},
Eloy de Lera Acedo\altaffilmark{10},
Joshua S. Dillon\altaffilmark{5,11},
Roger Dickenson\altaffilmark{3},
Phillip Doolittle\altaffilmark{3},
Dennis Egan\altaffilmark{3},
Mike Hedrick\altaffilmark{3},
Patricia Klima\altaffilmark{3},
Saul Kohn\altaffilmark{6},
Patrick Schaffner\altaffilmark{3},
John Shelton\altaffilmark{3},
Benjamin Saliwanchik\altaffilmark{12},
H.A. Taylor\altaffilmark{3},
Rusty Taylor\altaffilmark{3},
Max Tegmark\altaffilmark{1,2},
Butch Wirt\altaffilmark{3},
}

\affil{\altaffilmark{1}MIT Kavli Institute for Cosmological Physics, Cambridge, MA, 02139 USA}
\affil{\altaffilmark{2}MIT Dept. of Physics, Cambridge, MA, 02139 USA}
\affil{\altaffilmark{3}National Radio Astronomy Obs., Charlottesville, VA}
\affil{\altaffilmark{4}Dept. of Electrical and Computer Engineering, University of Virginia, Charlottesville, VA 22904, USA}
\affil{\altaffilmark{5}Dept. of Astronomy, University of California, Berkeley, CA, USA}
\affil{\altaffilmark{6}Dept. of Physics and Astronomy, University of Pennsylvania, Philadelphia, PA, USA}
\affil{\altaffilmark{7}Arizona State University, School of Earth and Space Exploration, Tempe, AZ 85287, USA}
\affil{\altaffilmark{8}Dept. of Electrical and Electronic Engineering, Stellenbosch University, Stellenbosch, SA}
\affil{\altaffilmark{9}SKA South Africa, Cape Town, SA}
\affil{\altaffilmark{10}Cavendish Laboratory, University of Cambridge, Cambridge, UK}
\affil{\altaffilmark{11}Berkeley Center for Cosmological Physics, Berkeley, CA, USA}
\affil{\altaffilmark{12}Astrophysics and Cosmology Research Unit, University of KwaZulu-Natal, Durban, SA}
\begin{abstract}
We use time-domain electromagnetic simulations to determine the spectral characteristics of the Hydrogen Epoch of Reionization Arrays (HERA) antenna. These simulations are part of a multi-faceted campaign to determine the effectiveness of the dish's design for obtaining a detection of redshifted 21\,cm emission from the epoch of reionization. Our simulations show the existence of reflections between HERA's suspended feed and its parabolic dish reflector that fall below -40\,dB at 150\,ns and, for reasonable impedance matches, have a negligible impact on HERA's ability to constrain EoR parameters. It follows that despite the reflections they introduce, dishes are effective for increasing the sensitivity of EoR experiments at relatively low cost. We find that electromagnetic resonances in the HERA feed's cylindrical skirt, which is intended to reduce cross coupling and beam ellipticity, introduces significant power at large delays ($-40$\,dB at 200\,ns) which can lead to some loss of measurable Fourier modes and a modest reduction in sensitivity. Even in the presence of this structure, we find that the spectral response of the antenna is sufficiently smooth for delay filtering to contain foreground emission at line-of-sight wave numbers below $k_\parallel \lesssim 0.2$\,$h$Mpc$^{-1}$, in the region where the current PAPER experiment operates. Incorporating these results into a Fisher Matrix analysis, we find that the spectral structure observed in our simulations has only a small effect on the tight constraints HERA can achieve on parameters associated with the astrophysics of reionization.
\end{abstract}
\section{Introduction}
Observations of the redshifted 21\,cm radiation from neutral hydrogen in the intergalactic medium (IGM) have the potential to illuminate the hitherto unobserved {\it dark ages} and {\it cosmic dawn}, revolutionizing our understanding of the first UV and X-ray sources in the universe and how they influenced galactic evolution (see \citet{Furlanetto:2006Review}, \citet{Morales:2010}, and \citet{Pritchard:2012} for reviews). Major experimental endeavors are underway to detect the 21\,cm signal, with most focusing on the epoch of reionization (EoR) during which UV photons from early galaxies converted the hydrogen in the universe from neutral to ionized. One approach involves measuring the sky-averaged global signal and is being pursued by experiments such as EDGES \citep{Bowman:2010}, LEDA \citep{Greenhill:2012,Bernardi:2015}, DARE \citep{Burns:2012}, SciHi \citep{Voytek:2014}, ZEBRA \citep{Ekers:2012}, SARAS \citep{Patra:2015}, and BIGHORNS \citep{Sokolowski:2015} either in their planning stages or already taking data. The global signal is also potentially observable with a zero-spacing interferometer \citep{Presley:2015,Singh:2015,Venumadhav:2015}. Another strategy is to observe spatial  fluctuations in the 21\,cm emission using radio interferometers. A first generation of such experiments are attempting to obtain upper limits or a first statistical detection of the power spectrum of 21\,cm brightness temperature fluctuations. These include the Giant Metrewave Telescope (GMRT)  \citep{Paciga:2013}, the Low Frequency ARray (LOFAR), \citep{VanHaarlem:2013}, the Murchison Widefield Array (MWA) \citep{Tingay:2013a}, the MIT Epoch of Reionization Experiment (MITEoR) \citep{Zheng:2013}, and the Precision Array for Probing the Epoch of Reionization (PAPER) \citep{Parsons:2010}. Already, many of these experiments are beginning to yield upper limits on the 21\,cm signal \citep{Dillon:2013,Parsons:2014,Jacobs:2015,Dillon:2015b,Trott:2016,EwallWice:2015a} and significant scientific results. The most stringent power-spectrum upper limit of $\approx 500$\,mK$^2$ by PAPER \citep{Ali:2015} is able to rule out a number of scenarios in which the intergalactic medium received little or no heating from X-rays \citep{Pober:2015,Greig:2015b}. As these current observatories reach their sensitivity limits, work is beginning on the next generation of interferometers, which will have the sensitivity required for a robust detection of the 21\,cm power spectrum. These include the Square Kilometer Array (SKA-1 Low) \citep{Mellema:2013} and the Hydrogen Epoch of Reionization Array (HERA)\citep{Pober:2014,DeBoer:2016,Dillon:2016} .   

The primary challenge to obtaining a high redshift detection of the cosmological signal through both of these methods is the existence of foregrounds that are $\approx 10^5$ times brighter \citep{Bernardi:2009,Pober:2013a,Dillon:2014}. Fortunately, it is expected that the foregrounds are spectrally smooth. In power-spectrum experiments, smooth foregrounds are naturally contained to a finite region of Fourier space, corresponding to large line of sight scales, known as the {\it wedge} \citep{Datta:2010,Vedantham:2012,Parsons:2012b,Thyagarajan:2013,Liu:2014a,Liu:2014b}.

 Since the location of each foreground on the sky determines its position in the wedge, with sources near the horizon being introduced at line of sight scales closest to the {\it EoR window}, the angular response of an instrument has significant implications on the amount of side-lobe power that is leaked into the EoR window. In \citet{Thyagarajan:2015a,Thyagarajan:2015b} and \citet{Pober:2016}, it is found that the response of the antenna, near the horizon, has a significant impact on foreground contamination and that centralized beams with highly suppressed sidelobes are preferable.   

Beyond leakage from wedge sidelobes, any structure in the frequency response of the instrument is imprinted on the foregrounds and has the potential to leak power into the EoR window at small line of sight scales, masking the signal. Indeed, sub-percent spectral features in the analogue and digital signal chains on the initial build-out of the MWA are proving to be a significant calibration challenge \citep{Dillon:2015b,EwallWice:2015a,Beardsley:2016}. 

In principle, spectral structure in the bandpass of an instrument may be removed in calibration. However, many approaches rely on detailed models of the foregrounds themselves and imperfections in these models may introduce spurious spectral structure (e.g. \citet{Barry:2016,Patil:2016}; Ewall-Wice et al. in preparation). Redundant calibration \citep{Wieringa:1992,Liu:2010,Zheng:2014} avoids relying on a detailed sky models but in its current state would be unable to correct direction dependent chromaticity that varies from antenna to antenna. Furthermore, direction dependent calibration requires imaging which is not performed in the delay power spectrum technique \citep{Parsons:2012b} that HERA is designed for.  Because of our limited knowledge of low frequency foregrounds and the fidelity of calibration algorithms, it is important to design experiments whose bandpasses are as devoid of spectral structure as possible.

\begin{figure}
\includegraphics[width=.5\textwidth]{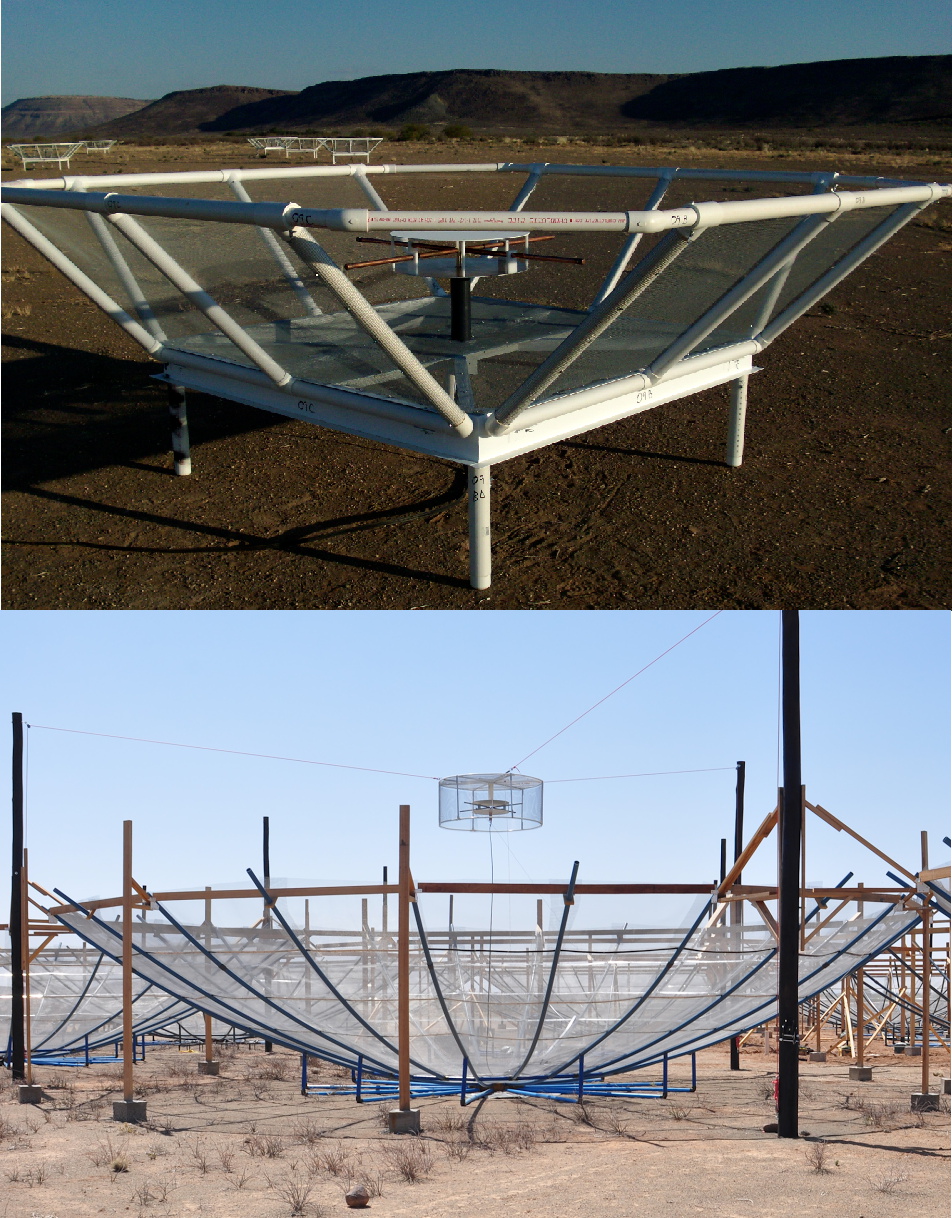}
\caption{The HERA antenna element (bottom) uses a parabolic dish to achieve an order of magnitude increase in collecting area over the PAPER antenna (top). The sleeved dipole in the center of the PAPER backplane is identical to the sleeved dipole being suspended under the cylindrical skirt over the vertex of the HERA dish. The suspended feed arrangement has the potential to introduce intra-antenna reflections which we explore in this work.}
\label{fig:AntennaCompare}
\end{figure}

Drawing from the lessons of the PAPER, MWA, and MITEoR experiments, HERA \citep{Pober:2014,DeBoer:2016,Dillon:2016} is a next generation 21\,cm experiment designed to achieve a two-orders of magnitude improvement in sensitivity over current efforts, allowing it to make a robust detection of the 21\,cm power spectrum during the EoR. Much of this sensitivity increase is enabled by moving the collecting area of the instrument into short baselines, with more modes outside of the wedge, and a switch from PAPER's skirted dipoles and the MWA's phased dipole arrays to an antenna element that consists of a feed suspended over a large reflecting parabolic dish. In Fig.~\ref{fig:AntennaCompare} we show a PAPER antenna and one of the initial 19 HERA dishes currently being deployed in South Africa at the same site. A central requirement for HERA's dish design is that the antenna have a response that is sufficiently smooth in frequency and a narrow beam, leaving the EoR window free of supra-horizon emission.

This paper and its companions \citep{Neben:2016,Patra:2016,Thyagarajan:2016} describe a multi-pronged campaign to assess the requirements and performance of the HERA dish for isolating foregrounds within the wedge. We accomplish this by establishing antenna specifications with simulations of foregrounds using the Precision Radio Interferometry Simulator (PRISim\footnote{\url{https://github.com/nithyanandan/PRISim}})\citep{Thyagarajan:2016} and verifying that the HERA primary antenna element meets these specifications with reflectometry \citep{Patra:2016} and ORBCOMM beam mapping \citep{Neben:2016}. In this work, we present the results of time-domain electromagnetic simulations that are intended to predict the degree of spectral structure due to reflections in the HERA dish, assess the impact of this spectral structure on the leakage of foregrounds into the EoR window, and to verify consistency with the reflectometry measurements taken in the field. 

This paper is organized as follows. In \S~\ref{sec:Formalism} we lay out our analytic framework, describing the impact of signal-path delays (from reflections or otherwise) on foreground leakage in delay-transform power spectra. In \S~\ref{sec:Simulations} we describe our time-domain electromagnetic simulations of the HERA dish element and how we extract the voltage response function. In \S~\ref{sec:Results} we describe the results and their origin in the antenna geometry. We also verify our simulation framework by comparing its prediction for the $S_{11}$ reflection coefficient of the HERA dish to direct field measurements described in \citet{Patra:2016}. In \S~\ref{sec:Sensitivity} we apply our electromagnetic simulation results to a foreground model to determine the extent that the HERA dish's chromatic structure compromises the EoR window and the impact on HERA's ability to constrain reionization parameters. We summarize our conclusions in \S~\ref{sec:Conclusion}.

\section{The Impact of Reflections on Delay-Transform Power Spectra}\label{sec:Formalism}
In this section, we show how delayed signals in the analog signal path of an antenna lead to foreground contamination of the EoR window. Intuitively, any reflections in the signal path introduce frequency ripples in the gain of the instrument. Since time delay is the Fourier dual to frequency \citep{Parsons:2012b}, reflections with larger delays introduce ripples with shorter periods. Any high-delay frequency structure imprinted on the much brighter foregrounds has the potential to mimic and swamp the signal unless it is brought below the ratio  between the foregrounds and the weak cosmological signal. Equations describing the effect of direction independent reflections in an interferometers signal chain downstream of the feed are derived in \citet{EwallWice:2015a}. We now extend this analysis by considering the direction dependent reflections that can occur within the antenna element. 
We start by denoting the electric field of radiation arriving from direction $\bl$ at the $i^{th}$ antenna element at position $\bx_i$ with angular frequency $\omega$ as $s(\bl,\bx_i,\omega)$ and the time-domain field as $\fts(\bl,\bx_i,t)$. Using the fact that $\widetilde{s}(\bl,\bx_i,t)=\widetilde{s}(\bl,{\bf 0},t-\tau_i) \equiv \widetilde{s}(\bl, t-\tau_i)$, where $\tau_i=\bl \cdot \bx_i /c$, we omit $\bx_i$ from our notation, referring to $\widetilde{s}(\bl,\bx_i,t)$ as $\widetilde{s}(\bl,t-\tau_i)$.

Reflections within the signal chain of each antenna are generally described by a complex direction-dependent reflection coefficient, $r_i(\bl,\omega)$ which we refer to as $\ftr_i(\bl,\tau)$ in the delay domain. The effect of each reflection with delay $\tau$ is to add the signal to itself multiplied by $\ftr_i$ and delayed by $\tau$. The voltage signal measured at the $i^{th}$ antenna element, $\widetilde{v}_i$, is the integral over solid angle of the electric fields arriving from all directions. The presence of reflections introduces a convolution of the electric field entering the antenna (delayed by $\tau_i$) with $\ftr_i(\bl,\tau)$:

\begin{equation}\label{eq:Voltage}
\ftv_i(t) = \int d \Omega \int d \tau \, \ftr_i(\bl,\tau) \fts(\bl,t- \tau_i - \tau).
\end{equation}
Applying the Fourier convolution theorem, the Fourier transform of this equation gives $v_i(\omega)$ as the simple angular integral of the product of $s_i(\bl,\omega)$ and $r_i(\bl,\omega)$.
\begin{equation}
v_i(\omega) = \int d \Omega  \, r_i(\bl,\omega) s(\bl,\omega) e^{-i \omega \tau_i}.
\end{equation}

The correlator of a radio interferometer records the time-averaged product of the Fourier transfored voltage streams of the $i^{th}$ and $j^{th}$ antennas. 
The time averaged correlation between the two antennas is
\begin{align}\label{eq:Visibility}
V_{ij}'(\omega) &= \left \langle v_i(\omega) v_j^*(\omega) \right \rangle_t \nonumber \\
 &= \int d \Omega \, r_i(\bl,\omega) r_j^*(\bl,\omega) I(\bl, \omega) e^{- i\omega\Delta \tau_{ij}} \nonumber \\
 & = \int d \Omega \, r_i(\bl,\omega) r^*_j(\bl,\omega)  I(\bl,\omega)e^{-i\omega \bb_{ij} \cdot \bl/c},
\end{align}
where $\Delta \tau_{ij} = \tau_i-\tau_j = (\bx_i-\bx_j) \cdot \bl/c$, $I(\bl,\omega)=\left \langle |s(\bl,\omega)|^2 \right \rangle_t$ is the intensity, $\bb_{ij} = (\bx_i-\bx_j)$, and the 'prime' on $V_{ij}$ indicates that this is the measured visibility with antenna gains applied rather than the intrinsic visibility of the sky. Here, we have invoked the fact that electromagnetic waves arriving from different directions are incoherent \citep{Thompson:1986}. This equation is the familiar interferometry equation \citep{Thompson:1986}, allowing us to see that $r_i(\bl,\omega)$ is mathematically equivalent to the voltage beam of the antenna. This can be explained more intuitively by the fact that any component of the instrument that multiplies the signal in the frequency domain will convolve the signal in the delay domain. Convolutions of the signal in the time-domain are mathematically identical to the effect of reflections. Setting a specification on reflections is hence equivalent to setting a specification on the spectral smoothness of the antenna's voltage beam. 

In order to filter spectrally smooth foregrounds from the signal, many experiments are employing the {\it delay transform} over frequency, defined as \citep{Parsons:2012b}
\begin{equation}
\widetilde{V}_{ij}(\tau) = \frac{1}{2 \pi} \int d \omega V_{ij}(\omega) e^{i \omega \tau }.
\end{equation}
Applying this to equation~\ref{eq:Visibility}, we obtain
\begin{equation}\label{eq:DelayVis}
\widetilde{V}_{ij}(\tau)= \frac{1}{2\pi} \int d \Omega \int d \omega r_i(\bl,\omega) r_j^*(\bl,\omega)  I(\bl,\omega) e^{-i \omega ( \bb_{ij} \cdot \bl /c - \tau)}.
\end{equation}
Let us examine the quantity within the angular integral. Assuming that the voltage beam is perfectly flat in its spectrum and ignoring the smooth spectral structure in $I(\bl,\omega)$ we see that for fixed $\bl$, the $\omega$ integral will result in a dirac-delta function in delay, centered at $\tau =  \bb_{ij}\cdot \bl/c$. hence each source located at $\bl$ on the sky is mapped to a line $\tau =\bb_{ij}\cdot \bl/c$, resulting in the much discussed ``wedge" \citep{Datta:2010,Vedantham:2012,Parsons:2012b,Morales:2012,Thyagarajan:2013,Liu:2014a,Liu:2014b}.

\begin{figure*}
\includegraphics[width=\textwidth]{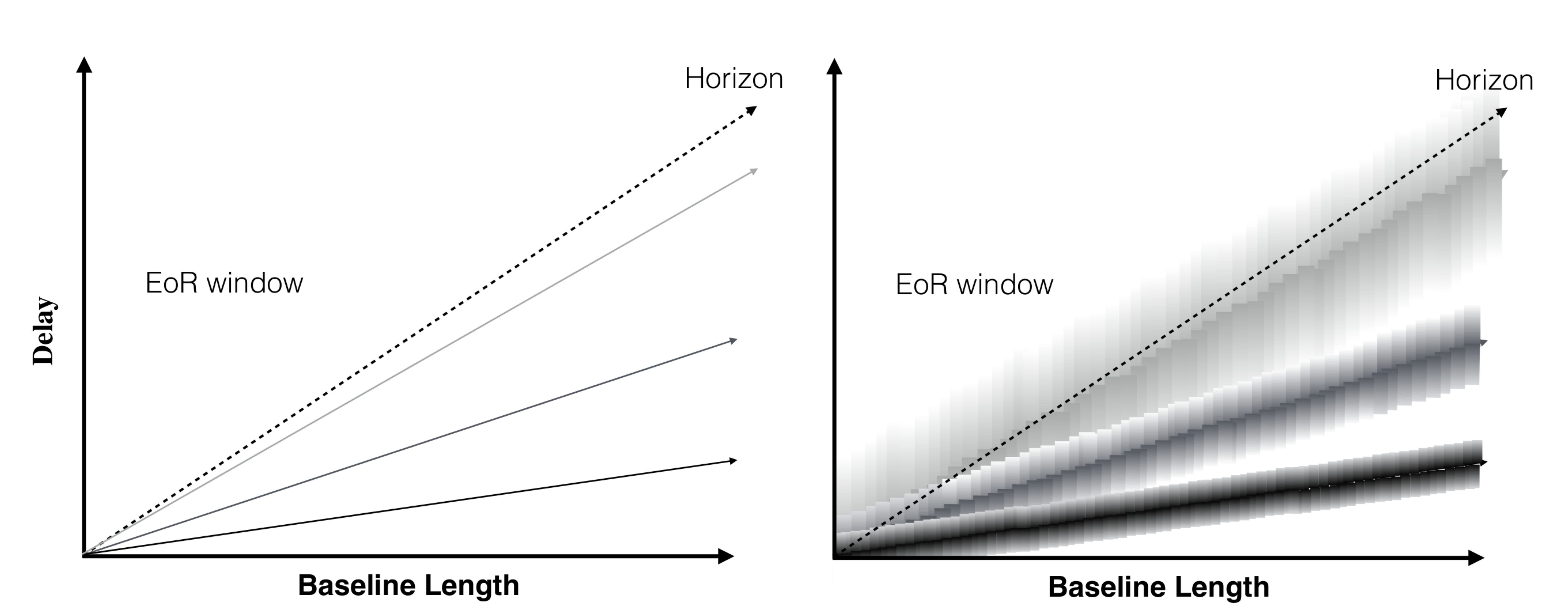}
\caption{A cartoon demonstration of the impact on foregrounds of the frequency dependent beam. Left: The location of three, spectrally flat, sources in delay space assuming a frequency independent beam (no reflections in the antenna element). Right: the presence of chromaticity due to delayed signal within the antenna smears the source in delay with the kernel given by equation~\ref{eq:Kernel}. Since the frequency response of the dish is a function of direction on the sky, the shape of the delay kernel is different for each source line. We see that this smearing can lead to substantial supra-horizon emission. In this paper, we consider a direction independent delay-kernel that is source primarily by reflections while a more general direction-dependent kernel is explored in \citet{Thyagarajan:2016}}
\label{fig:Smearing}
\end{figure*}

 The presence of a realistic frequency dependent beam causes each source line to be convolved in delay with the direction dependent kernel. To see this, we carry out the $\omega$ integral in equation \ref{eq:DelayVis} and apply the Fourier Convolution theorem. 
 \begin{equation}
 \widetilde{V}_{ij}(\tau) = \int d \Omega \int d \tau' \widetilde{I}(\bl,\bb_{ij} \cdot \bl/c - \tau-\tau') \widetilde{R}(\bl,\tau'),
 \end{equation}
 where
\begin{equation}\label{eq:Kernel}
\ftR_{ij}(\bl,\tau) = \int d  \tau' \ftr_i(\bl,\tau'-\tau) \ftr^*_j(\bl,\tau').
\end{equation}
Thus, each source line in the wedge is convolved with an antenna kernel, $\ftR_{ij}$, which is itself the convolution of the time-reversed delay response of the voltage beam of antenna $i$ with the complex conjugate of voltage beam of antenna $j$. In the remainder of this paper, we will often refer to $\ftR_{ij}$ as the {\it power kernel} applied to a visibility. We demonstrate the effect of foreground smearing in Fig.~\ref{fig:Smearing} for a simple model with only three sources. Without reflections, the sources would form lines intersecting zero in the two dimensional space with the baseline length, $b$ on the x-axis and the delay, $\tau$ on the y-axis. With the reflections, the sources are smeared out, leading to supra-horizon contamination. 
For the sake of simplicity, we now consider the case where the beam can be separated into independent angular and frequency components, $r_i(\bl,\omega) = g_i(\omega)a_i(\bl)$. For such a case, every line in Fig.~\ref{fig:Smearing} would be convolved with the same delay dependent shape, normalized to the gain of $a_i(\bl)$. In this situation, we have
\begin{equation}
\widetilde{V}_{ij}(\tau) = \int d\tau' \int d \tau'' \widetilde{g}_i(\tau' - \tau'')\widetilde{g}^*_j(\tau'') \widetilde{V}_{ij}^a(\tau-\tau'),
\end{equation}
where $V_{ij}^a$ is the visibility for the achromatic voltage pattern, $a_i(\bl)$. 
We can gain further insight into the behavior of the delay kernel arising from chromaticity by assuming that $\widetilde{g}_i(\tau=0)\gg\widetilde{g}_i(\tau>0)$, which should be the case at large delays for the smooth bandpasses our instruments are designed to have. 
\begin{align}\label{eq:KernelApprox}
\widetilde{V}_{ij}(\tau) &\approx \widetilde{g}_i(0)\int d \tau' \widetilde{g}_j^*(\tau')\widetilde{V}^a_{ij}(\tau- \tau') \nonumber \\
& + \widetilde{g}_j^*(0) \int d \tau' \widetilde{g}_i(-\tau')\widetilde{V}^a_{ij}(\tau-\tau').
\end{align}
Hence, to first order, the impact of reflections is to convolve the delay-transformed visibility with the voltage beam of the instrument. This may be a somewhat counter-intuitive result since naive dimensional analysis might predict that the power-kernel is proportional to the square of the delay-response. This linear relation requires that the voltage beam fall roughly five orders of magnitude (the same as the dynamic range between foregrounds and signal) in the regions of delay space that we want to measure the signal.

\section{Electromagnetic Simulations of the HERA dish element}\label{sec:Simulations}

Having formally derived the impact of reflections on foreground visibilities, we are in a position to investigate their existence in the HERA dish. In this paper, we assume a separable beam kernel whose response is given at zenith and whose high-delay components are sourced by intra-dish reflections (equation~\ref{eq:KernelApprox}). The impact of the direction dependent kernel which includes side-lobe chromaticity is investigated in \citep{Thyagarajan:2016}. In this section, we describe the setup and parameters of our simulations (\S~\ref{ssec:Simulations}), and how we extract the voltage response function of the dish (\S~\ref{ssec:Deconvolve}).

\subsection{The Simulations}\label{ssec:Simulations}
 The time-dependent electromagnetic behavior of the antenna was modeled using Microwave Studio, a commercial numerical simulation package produced by Computer Simulation Technology (CST).  The model consists of an idealized 14 m diameter paraboloid reflector with a feed structure placed at the focus (4.5 m above the surface). The feed is a dual linear polarized sleeved dipole identical to that used as an element of PAPER but with a cylindrical skirt used in place of the angled planar reflectors.   The dipole arms are modeled as copper tubing with aluminum disks and reflector surfaces.  For simulation, the entire model was encapsulated in vacuum dielectric with radiative boundaries and discrete ports of 125\,$\Omega$ impedance were defined at the terminals of the orthogonal dipoles. The simulation is conducted in a box encapsulating the dish geometry plus three wavelengths (at 100\,MHz) on all sides with open boundary conditions. 

Microwave Studio divides the model into approximately 94 million discrete hexahedral cells where the field is calculated over time using Finite Integration Technique incorporating the Perfect Boundary Approximation\footnote{This proprietary technique is used by CST to reduce the number of simulation cells.}  \citetext{Workflow and Solver Overflow Document, CST Microwave Studio, 2015, Chapter 3.} in response to a 100-200 MHz broadband pulse of about 40 ns in duration.  The model was excited in two ways: 1) via a E-W polarized plane wave entering the model along the bore sight, and 2) from one of the discrete terminals (whose analysis we focus on in \S~\ref{ssec:S11}).   The voltage responses at the terminals are monitored for these two cases over a duration of 500 ns as the excitation pulse travels throughout the model.

\begin{figure*}
\includegraphics[width=\textwidth]{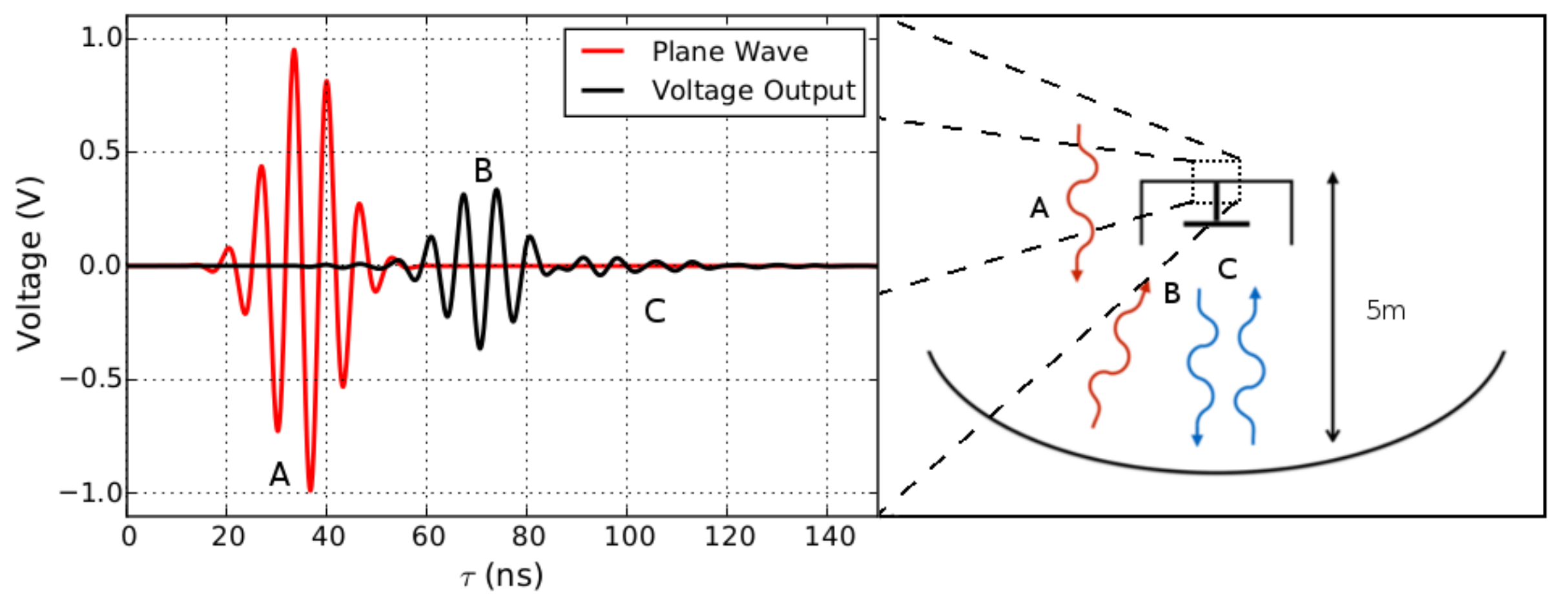}
\caption{An illustration of our simulation products and their origin in the HERA antenna geometry. A plane wave is injected from above the feed. The electric field of the plane wave at the feed terminals (red line) along with the voltage output is recorded (black line). The feed in our simulation is situated $5$\,m above the bottom of the dish, hence there is a $\approx 30$\,ns delay between when the plane wave passes the terminal for the first time (A) and when it is first absorbed in the dipole (B), leading to the voltage response. Of concern to 21\,cm experiments are the subsequent reflections between the feed and the dish (C) which can lead to larger delays that contaminate the EoR window.}
\label{fig:SimulationOutput}
\end{figure*}

In Fig.~\ref{fig:SimulationOutput} we show the geometry of the electromagnetic simulation. A $150$\,MHz plane wave with a Gaussian envelope is initialized above the dish vertex traveling in the $-z$ direction. It is reflected by the dish before entering the dipole feed, hidden below the cylindrical skirt in this figure. We record the electric field voltage of the plane wave at the feed output terminals as a function of time, plotted as a red line in Fig.~\ref{fig:SimulationOutput} along with the output voltage at the terminals for the feed polarized parallel to the plan wave (black line). The delay between the central envelope of the plane wave and the voltage output of the feed is $\approx 30$\,ns which corresponds to the round trip travel time from the feed to the dish vertex and back. However, while the input plane wave, modulated by a gaussian, falls off rapidly after the first $\approx 20$\,ns after its peak, we see that the voltage output decays far more slowly due to reflections and resonances within the antenna structure. We are able to get a qualitative feel for the amplitude of the reflections by inspecting the falloff of the time domain voltage response and see that after $60$\,ns it reaches $\approx -25$\,dB.

\subsection{Deconvolving the Response Function}\label{ssec:Deconvolve}

We can do much better than this. From equation~\ref{eq:Voltage}, we know that the voltage output results from the convolution of the plane wave input with the voltage gain of the antenna. Since we know the input wave, a straightforward application of the Fourier convolution theorem allows us to determine the voltage response.

\begin{figure}[h!]
\includegraphics[width=.5\textwidth]{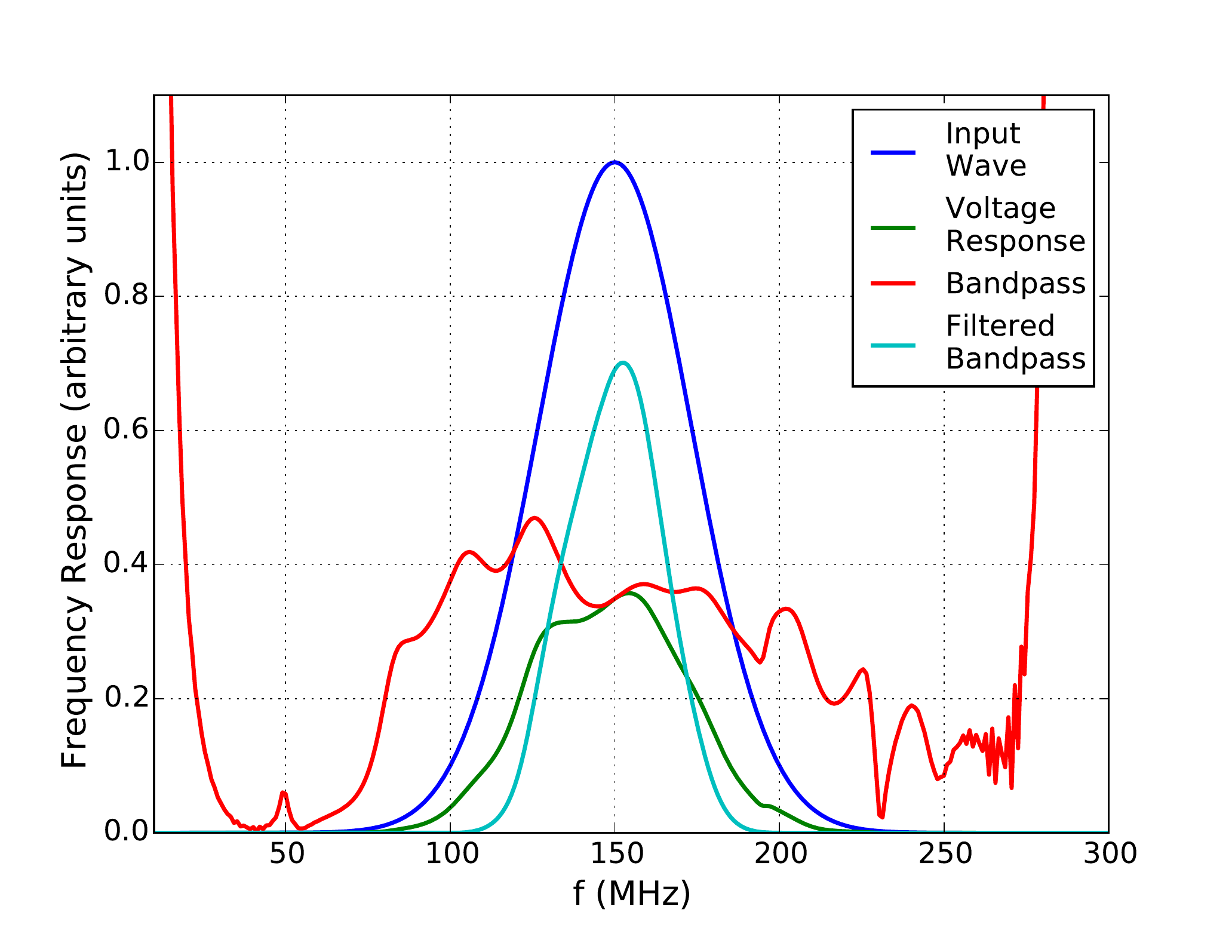}
\caption{The absolute value of the Fourier transform of the voltage output from our dish simulations (green line) and the input wave (blue line), normalized to the amplitude of the input wave at $150$\,MHz. The ratio between input and output is plotted as a red line. Since our input is limited to frequencies between $\approx 20$ and $280$\,MHz, there are significant numerical artifacts in the ratio that causes divergence towards the plot edges. To eliminate this noise, we multiply by a Blackman-Harris window between $100$ and $200$\,MHz and set our estimate to zero elsewhere (cyan line).}
\label{fig:FrequencyDomain}
\end{figure}

Since our simulation is sampled in finite time steps, we will adopt discretized notation for this section. In particular, our simulation consists of $N$ samples, evenly spaced by $d \tau$ at times $\tau_n = n \times d \tau$. 
We denote the output voltage at the feed terminals at time $\tau_n$ as $\widetilde{v}_n$. Rewriting the convolution in equation~\ref{eq:Voltage} in discrete notation, we have
\begin{equation}
\ftv_n(\bl) = \sum_m \ftr_m(\bl) \fts_{n-m}(\bl).
\end{equation}
We may undo this convolution by taking a discrete Fourier transform (DFT) of both ${\bf \widetilde{v}}$ and ${\bf \widetilde{s}}$ in time, dividing them in Fourier space, and taking an inverse DFT back. Symbolically,
\begin{equation}\label{eq:Inversion}
\ftr(\bl) = \boldsymbol{\mathcal{F}}^{-1} \left[ \frac{\boldsymbol{\mathcal{F}} {\bf \widetilde{v}}(\bl)}{\boldsymbol{\mathcal{F}}{\fts}(\bl)} \right] ,
\end{equation}
where $\boldsymbol{\mathcal{F}}$ is the Fourier transform matrix for a 1d vector of length $N$. 
\begin{equation}
\boldsymbol{\mathcal{F}}_{mn} = e^{-2 \pi i m n /N}.
\end{equation}
In Fig.~\ref{fig:FrequencyDomain} we show the amplitude of the Fourier transform of our Gaussian input, centered at $150$\,MHz along with the voltage response. Since our input only has support between $\approx 20$ and $280$\,MHz, the direct ratio of our voltage response and input wave is dominated by numerical noise outside of this range. We eliminate these artifacts by multiplying our ratio by a Blackman-Harris window between $100$\,MHz and $200$\,MHz and set our estimate to zero elsewhere. From a physical standpoint, this is sensible since 21\,cm experiments only observe a limited bandwidth. PAPER's correlator, which will initially serve as the HERA backend samples over a $100$\,MHz instantaneous frequency interval and analog filtering is applied to prevent aliasing.

\section{Simulation Results}\label{sec:Results}

We now discuss the results of our simulations. We focus on the time-domain die off of the voltage response and the resulting power kernel (\S~\ref{ssec:Kernel}), comparing it to an identical time-domain simulation of the skirted dipole antenna used by PAPER. In \S~\ref{ssec:Subbands} we investigate the dependence of the power kernel on frequency to determine whether specific parts of the HERA bandpass are more affected by reflections than others. In \S~\ref{ssec:Knee} we determine the origin of excess response at long delays in our simulations. In \S~\ref{ssec:Tradeoffs}, we discuss potential trade-offs in eliminating this structure from the antenna by determining the impact of removing components of the HERA feed that are responsible for the contamination. Finally, in \S~\ref{ssec:S11} we verify our simulation framework by comparing a separate time domain simulation of $S_{11}$ of the HERA dish to a direct field measurement with a vector network analyzer (VNA). 

\subsection{The Time Domain Response of the HERA Dish}\label{ssec:Kernel}
\begin{figure}[h!]
\includegraphics[width=.5\textwidth]{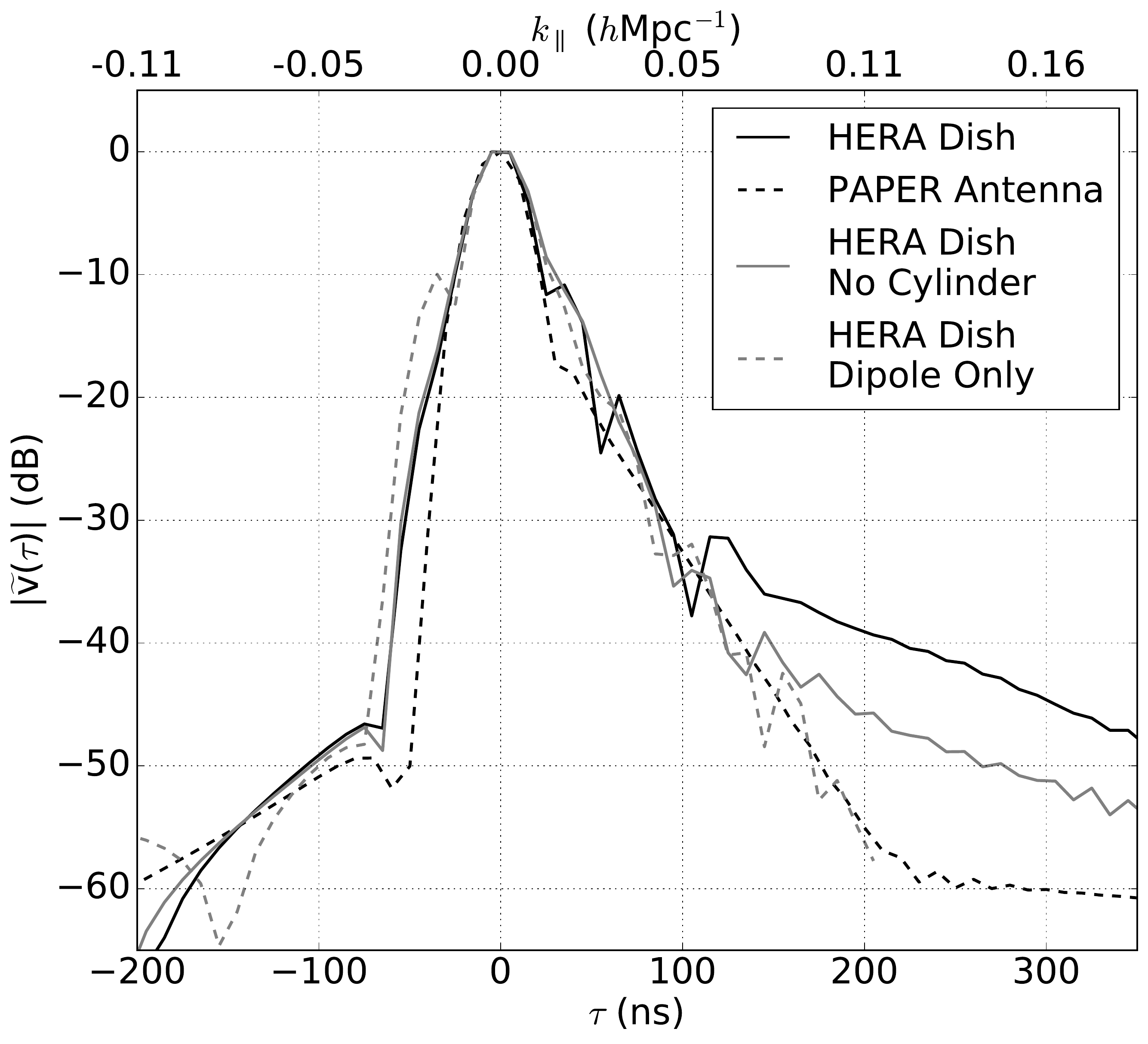}
\caption{The Fourier transform of the simulated voltage response of the HERA Dish (solid black line) and the PAPER antenna element (dashed black line). Reflections in the HERA dish element lead to significantly enhanced power above $\sim 50$\,ns. Since negative delays should be devoid of signal, they allow us to determine the dynamic range of our simulations which have a numerical noise/sidelobes floor of $-60$\,dB.}
\label{fig:SimulationResults}
\end{figure}

Applying equation~\ref{eq:Inversion} to our simulation, we obtain estimates of the time-domain voltage response of the HERA dish towards zenith which we plot in Fig.~\ref{fig:SimulationResults}. We also conduct a time-domain simulation of the voltage output in response to an identical input plane wave for the skirted dipole PAPER antenna (pictured above the HERA dish in Fig.~\ref{fig:AntennaCompare}) in order to determine whether the presence of the parabolic dish introduces reflections and spectral structure in excess of previous successful antenna designs.\footnote{To date, PAPER has produced the most stringent limits on the 21\,cm power spectrum leading us to use its design as a standard.}  We inspect the absolute value of $\ftv$ for both the PAPER and HERA antennas in Fig.~\ref{fig:SimulationResults}. Since non-zero values of $\ftv$ at negative delays violates causality, we assume such features are sourced by artifacts such as side-lobes of the zero-delay peak and/or numerical precision noise which sets a limit of $\approx -60$\,dB on the dynamic range of our method. We first note peaks in the HERA curve that are spaced by $\approx 35$\,ns and are absent in the PAPER simulation. Another significant difference between the two curves is a knee in the HERA gain at $\approx 120$\,ns that is not present in PAPER, leading to an increase in gain by approximately $20$\,dB at 200\,ns. Because this knee does not exhibit nodes at $35$\,ns intervals it is probably not sourced by reflections in the antenna-feed geometry but by reflections involving a geometric length that is not resolved by the 100\,MHz bandwidth of the simulation. 

\begin{figure}[h!]
\includegraphics[width=.5\textwidth]{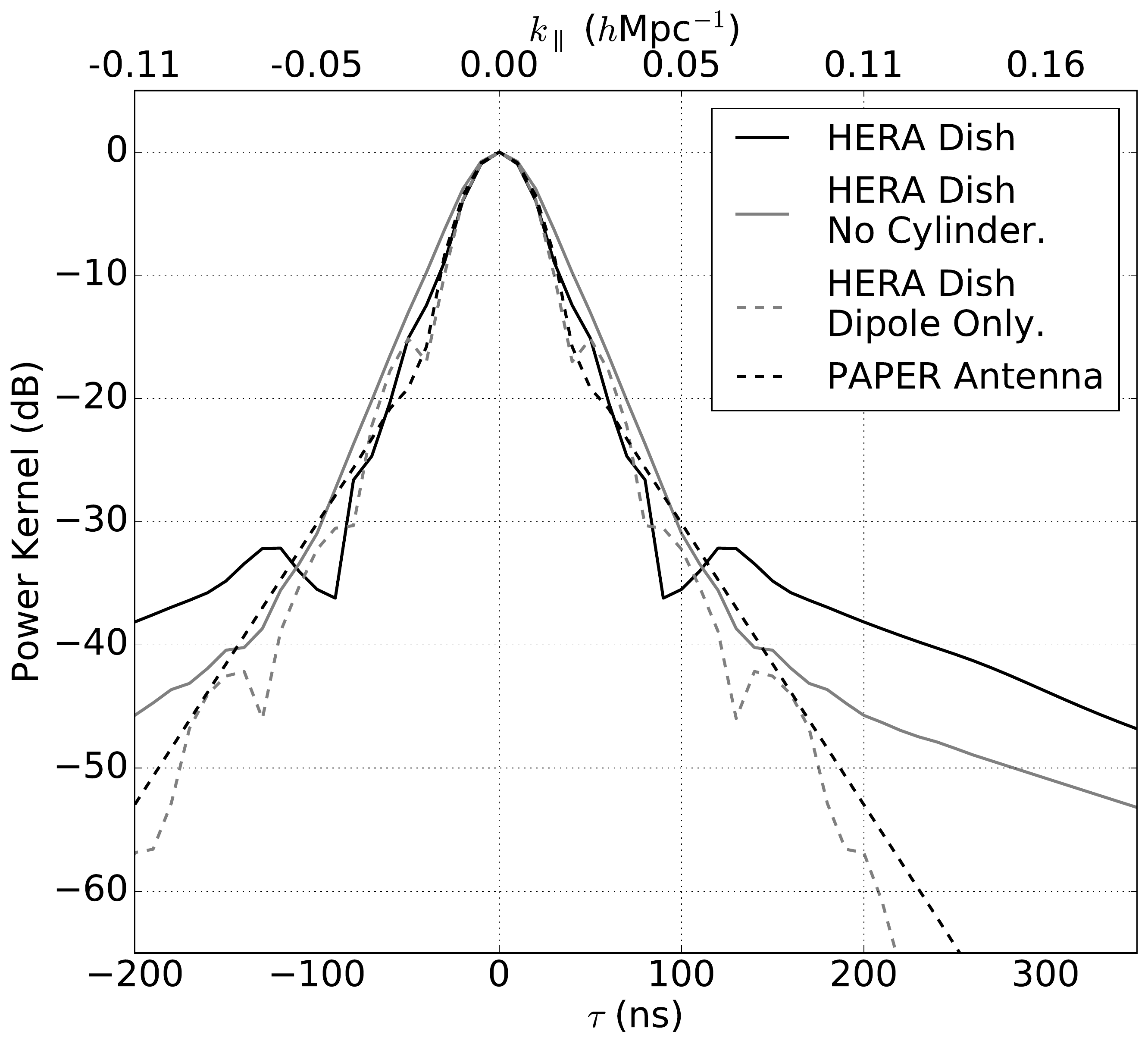}
\caption{The absolute value of the power kernel for the HERA dish (solid black line) and for the PAPER antenna element (dashed black line) calculated using equation~\ref{eq:Kernel}. While an antenna can only physically have a voltage response at positive delays, the delay kernel is formed from the convolution of one antenna with the time reversed conjugate response function of the other. Hence, the power kernel for two identical antennas will have $\widetilde{R}(\tau) = \widetilde{R}^*(-\tau)$.}
\label{fig:Kernels}
\end{figure}

Our next step is to compute the power kernel, $\widetilde{R}$, given by equation~\ref{eq:Kernel} by performing a convolution of the time-reversed voltage response with the complex conjugate voltage response. In Fig.~\ref{fig:Kernels} we show the resulting power kernel for PAPER and HERA. Since both voltage gains drop rapidly with increasing delay, the approximation in equation~\ref{eq:KernelApprox} holds quite well and we see that the kernels fall off at a rate similar to the voltage response. 

\subsection{The Delay Response of Subbands}\label{ssec:Subbands}
\begin{figure}[h!]
\includegraphics[width=.5\textwidth]{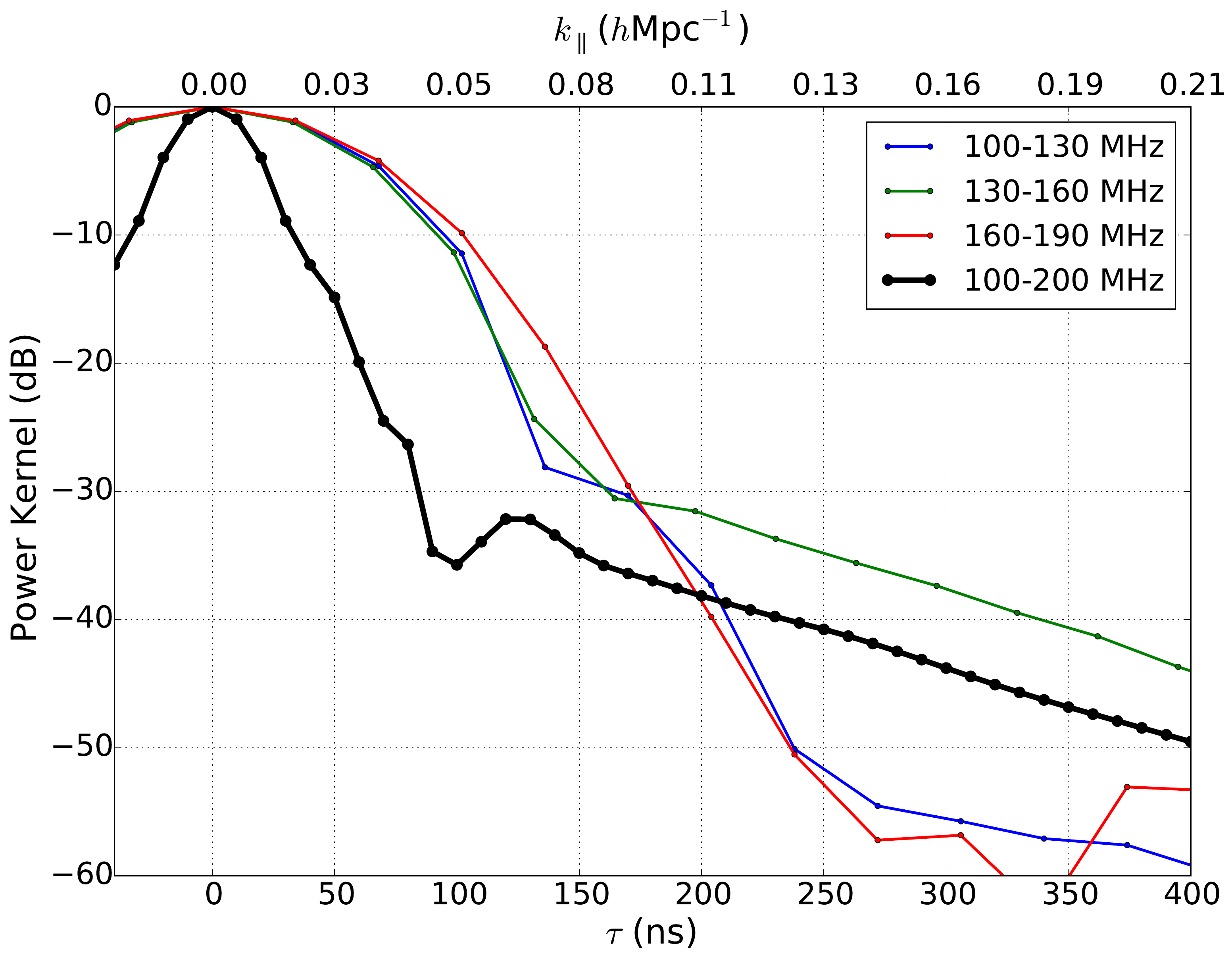}
\caption{The power kernel for the three subbands discussed in \S~\ref{ssec:Subbands} along with the kernel for the full bandwidth response function. While the long term falloff from reflections is prominent between $130-160$\,MHz, it appears at a much lower level in the other two subbands which fall below the central subband by $\sim 20$\,dB at $\sim 300$\,ns. $k_\parallel$ values for each delay are computed at $150$\,MHz. The wider central lobe below 150\,ns for the subband gains is due to the lower delay resolution from the smaller bandwidth. We also show the delay kernel for $100$\,MHz bandwidth (black thick line).}
\label{fig:KernelsSubbands}
\end{figure}

Because the 21\,cm brightness temperature fluctuations evolve over redshift intervals of $\Delta z \gtrsim 0.5$ \citep{Zaldarriaga:2004}, experiments will sub-divide their bands into $\approx 10$\,MHz intervals for power spectrum estimates at multiple redshifts. Thus the localization of reflections within the HERA band will determine whether all or only a subset of redshifts are affected. To determine whether the reflections are localized in frequency, we compute the voltage delay response and power kernel for three different subbands: $100-130$\,MHz, $130-160$\,MHz, and $160-190$\,MHz. In order to maintain decent resolution of the kernel itself, we use frequency ranges that are several times larger than the actual subbands that will be used for EoR power spectrum estimation ($\approx 10$\,MHz). The resulting power kernels are plotted in Fig.~\ref{fig:KernelsSubbands}. The central lobe of the subband kernels is significantly wider due to the the wider window functions incurred by the reduced bandwidth. We find that the shallow long-term falloff is only visible within the central $130-160$\,MHz band, indicating that long term reflections are isolated around $150$\,MHz.

\begin{figure}
\includegraphics[width=.5\textwidth]{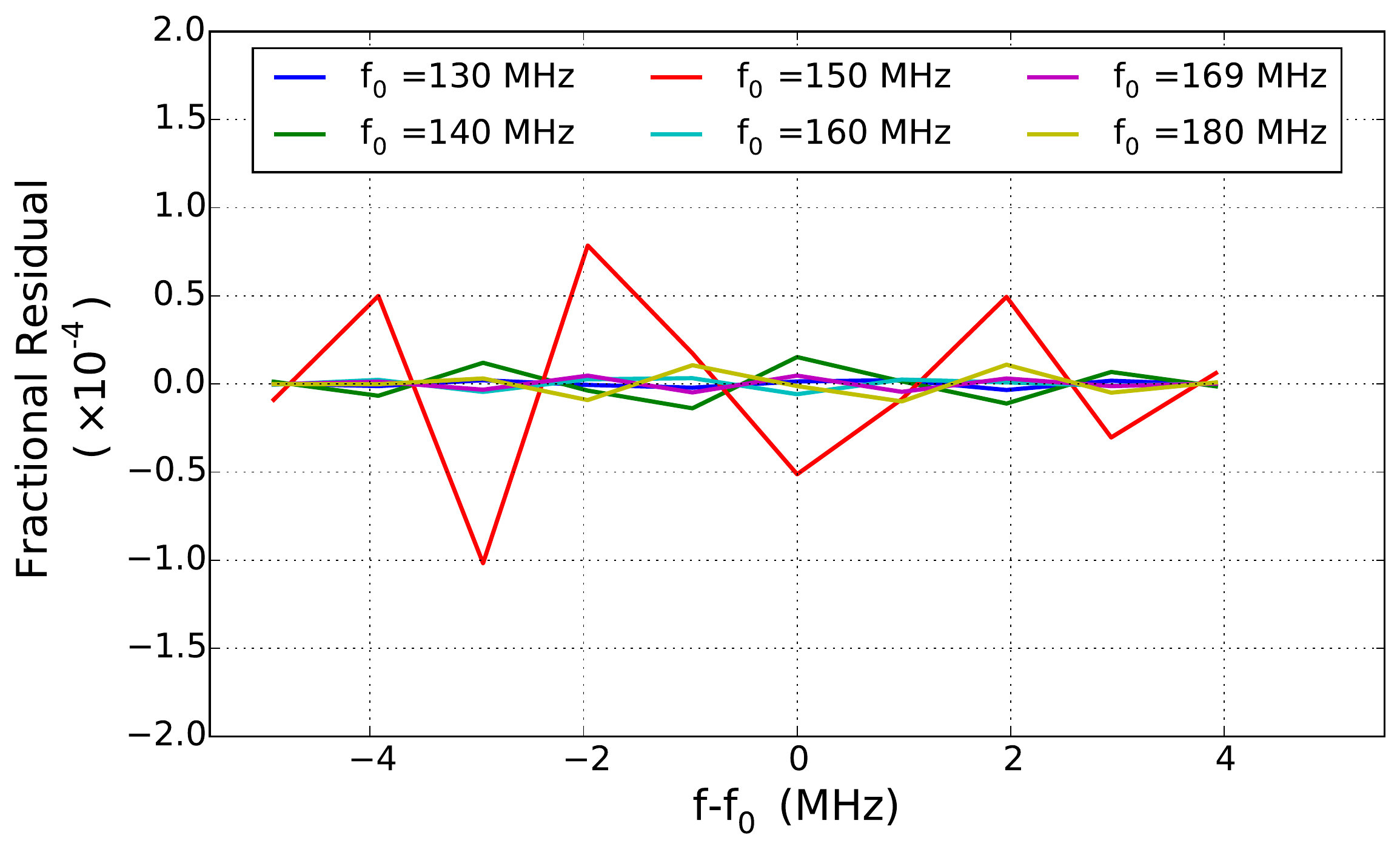}
\caption{Residuals on the absolute value of the gain over several subbands after fitting to a sixth order polynomial. Consistent with our findings in Fig.~\ref{fig:KernelsSubbands}, the fine frequency residuals in the 145-155\,MHz subband are over an order of magnitude greater than those in the other subbands.}
\label{fig:Residuals}
\end{figure}

To further illustrate the isolation of fine frequency structure in the center of the bandpass and to verify that our observations are not an artifact of our reduction of the simulation outputs, we fit $10$\,MHz intervals of the absolute value of the simulated gains to a sixth order polynomial and inspect the residuals in Fig.~\ref{fig:Residuals}. We find that the gain residuals on the sixth order fit are an order of magnitude greater on the $145-155$\,MHz subband than  any other frequency interval. 



\subsection{The Origin of the Knee.}\label{ssec:Knee}
In Fig.~\ref{fig:SimulationResults} we see that the long-term delay response of the HERA antenna differs from the PAPER design in two regards, the existence of node-like structures spaced every $\sim 35$\,ns, corresponding to the round trip delay between the dish and the feed, and a long-term knee that dominates the response function after 100\,ns but does not exibit any periodicity that might be associated with the feed-dish geometry. Here we establish the origins of the long time-scale structure using simulations. Reflectometry measurements conducted on the isolated feed in several different configurations \citep{Patra:2016} independently discover and verify these origins. To determine whether this knee-like structure originates from feed-dish reflections, we perform a simulation of a plane wave incident along the bore-sight the HERA feed without the dish and compare it with the time-domain response for the feed-and-dish system in Fig.~\ref{fig:FeedOnlyComparison}. The feed only simulation lacks the reflections associated with the dish and feed configuration (present at small delays) but retains the large-delay knee.  Thus, the dominant large time-scale contamination of the HERA band-pass is not intrinsic to the arrangement of a feed suspended over the dish.

\begin{figure}
\includegraphics[width=.4\textwidth]{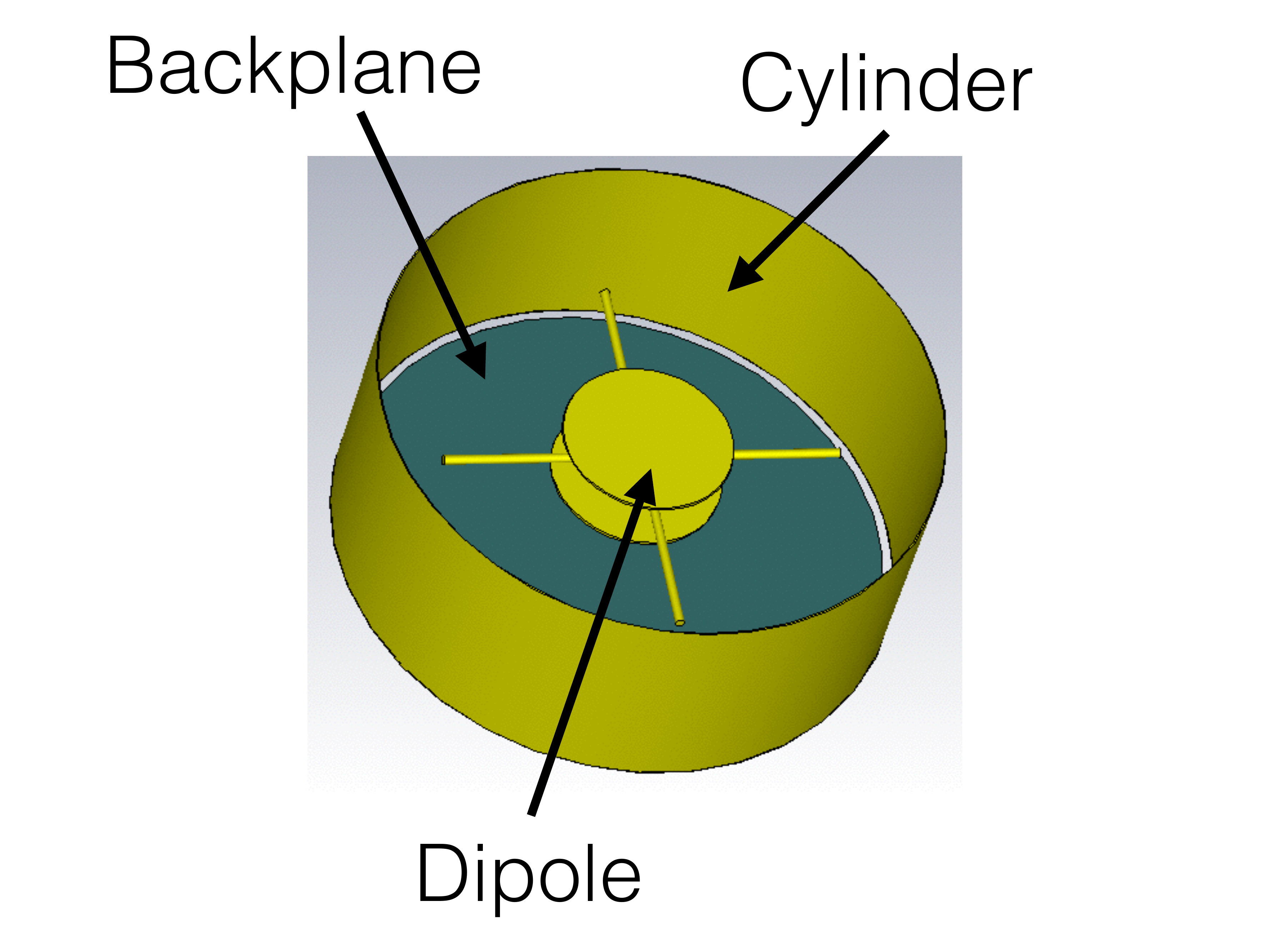}
\caption{A closeup rendering of the HERA feed which is suspended over the reflector, illustrating the cylindrical skirt, the backplane, and the dipole. Long time-scale spectral structure arises from electrical oscillations within the cylindrical cavity.}
\label{fig:FeedDetail}
\end{figure}

\begin{figure}
\includegraphics[width=0.5\textwidth]{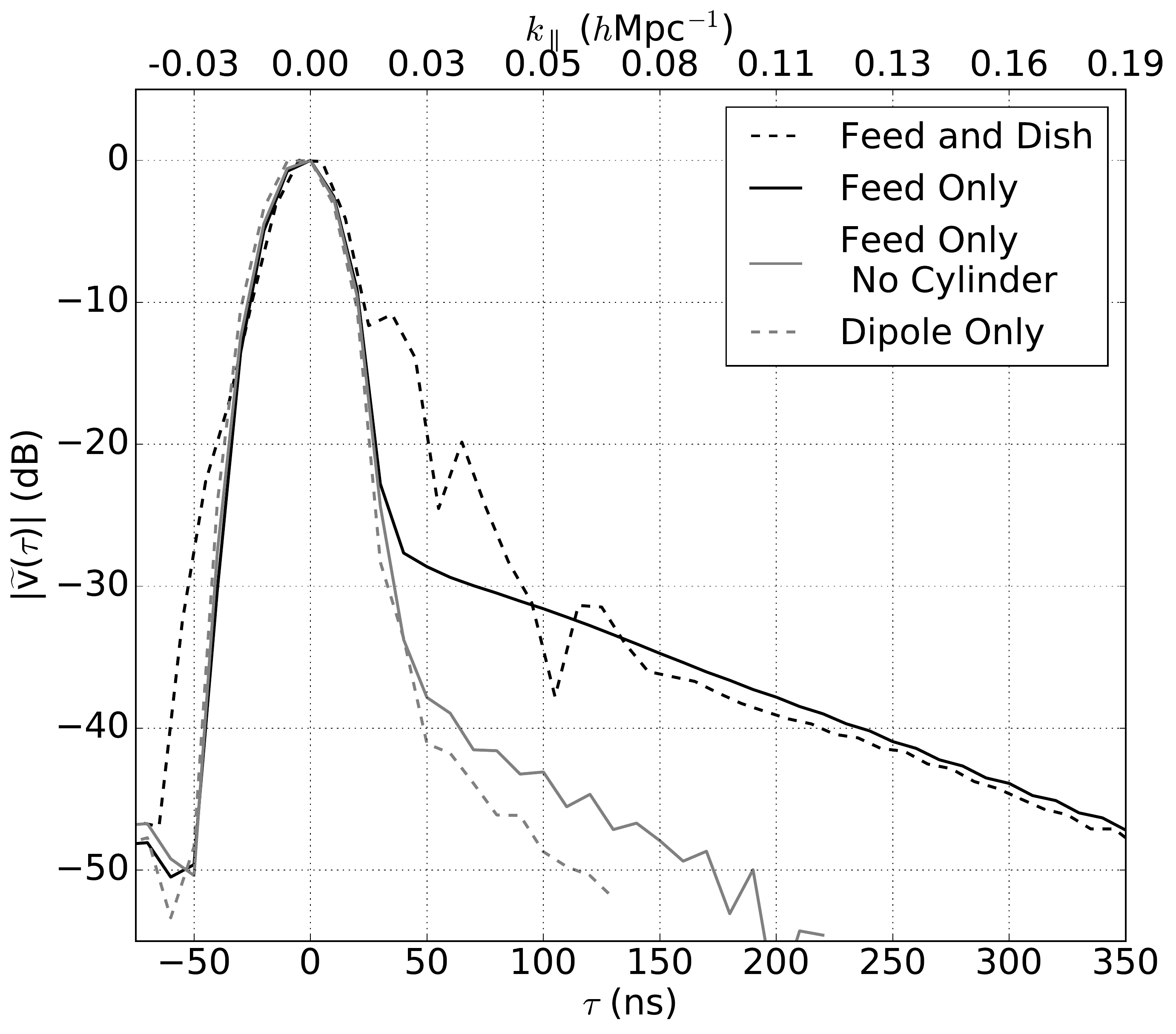}
\caption{The absolute value of the time-domain voltage response of the cylindrical dipole feed compared to the absolute value of the voltage response of the feed suspended over the dish. As we might expect, the $\sim 35$\,nm lobed structures associated with feed-dish reflections are absent from the simulation of the feed only. However, the knee like feature after $\sim 100$\,ns is. This indicates that the most severe spectral contamination in the current HERA design does not originate in reflections between the feed and the dish but rather within the feed itself.}
\label{fig:FeedOnlyComparison}
\end{figure}

We isolate the source of the knee within the feed structure (shown in Fig.~\ref{fig:FeedDetail}) by running simulations of the dipole feed with only a back-plane along with a sleeved dipole in isolation. We show both of these voltage responses in Fig.~\ref{fig:FeedOnlyComparison}. When the cylindrical skirt is removed from the feed, the knee vanishes. Because of the narrow band nature of the knee (\S~\ref{ssec:Subbands}), its presence only when the cylinder is attached, and its exponential falloff (characteristic of a damped harmonic oscillator), we conclude that the cylindrical skirt is behaving as a resonant cavity and the resulting stored oscillating fields are responsible for the majority of the fine-scale spectral structure in the HERA antenna's gain. We estimate the quality factor of the resonance to be $\approx 6.5$ from the time constant of the exponential falloff which yields a resonance width of $\approx 20$\,MHz. 

\begin{figure}
\includegraphics[width=.5\textwidth]{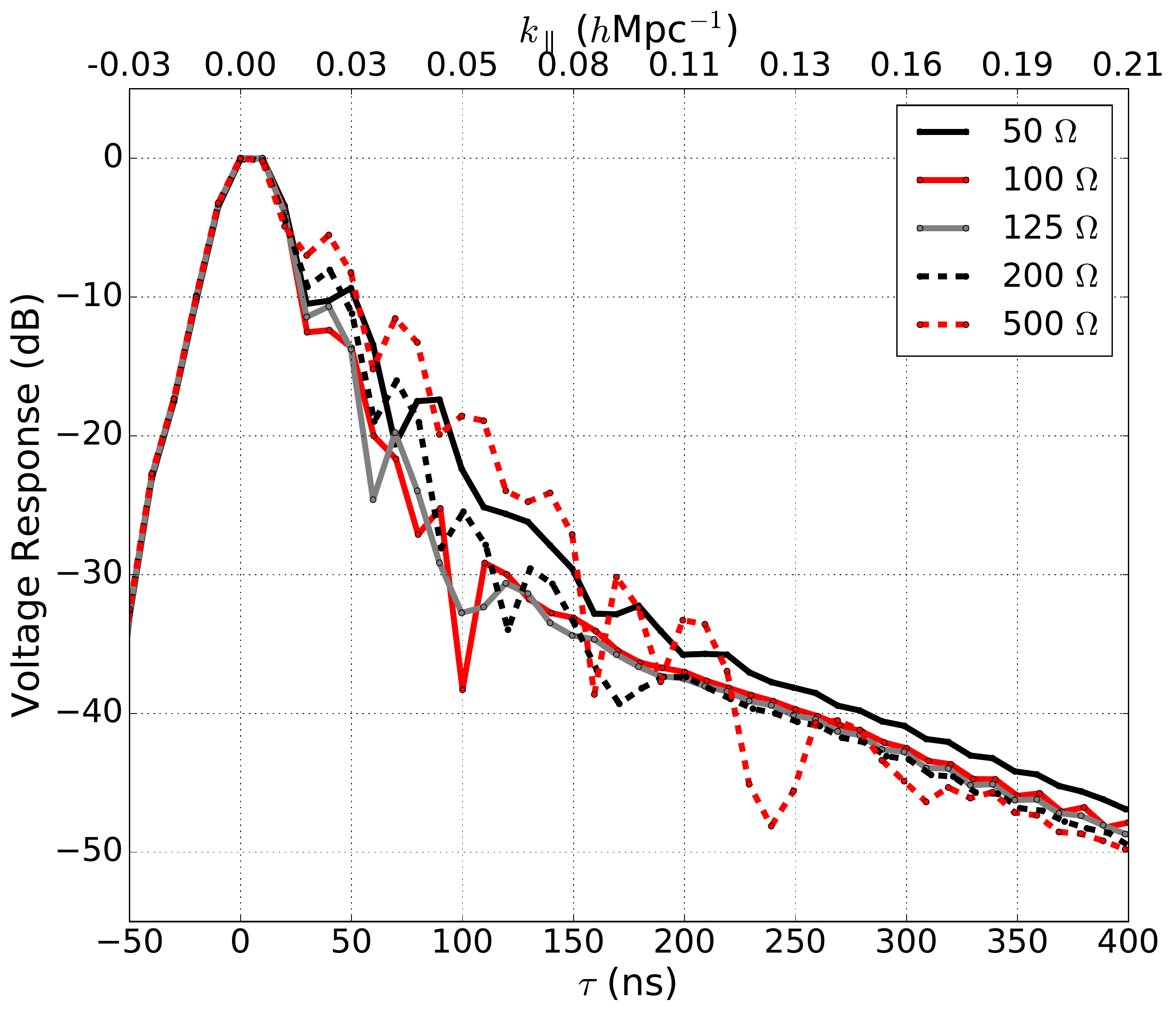}
\caption{The time-domain response of the HERA antenna towards a plane wave incident from zenith for a variety of termination impedances. As we vary the termination impedance, the structure which is dominated by feed-dish reflections, below $\approx 150$\,ns, varies significantly but leaves delays greater than $\approx 200$\,ns unchanged. Only in the extreme, 500\,$\Omega$ case do the reflections extend to large delays. Since structure above 250\,ns is primarily reponsible for contaminating the EoR window (\ref{fig:BothBaselines}), the termination impedance has a relatively small impact on HERA's overall sensitivity.  }\label{fig:ImpedanceCompare}
\end{figure}

Since the termination of the dipole could be significantly different from the termination used in the simulation $(100\Omega)$, we also investigate the impact of the termination impedance of the dipole on the delay-response. we vary the termination impedance between $50$ to $500\,\Omega$ for several plane-wave simulations and show the resulting voltage responses in Fig.~\ref{fig:ImpedanceCompare}. We find that changing the termination impedance has a significant impact on the time-domain response below $\approx 150$\,ns in the region that is dominated by dish-feed reflections. Since the termination impedance does not affect resonance of the feed cavity, its effect is small beyond $\approx 200$\,ns. Only in the extreme, 500\,$\Omega$ case do the reflections extend to appreciable delays. Since delays over $\approx 250$\,ns are responsible for contaminating the EoR window, termination impedance will only impact HERA's sensitivity for an extremely poor match.

Simulations combining the dish with the analog signal-chain show that matching networks can reduce the levels feed-dish reflections below those observed in this paper, even under realistic termination conditions \citep{Fagnoni:2016}\footnote{See Fig.~4. This paper investigates a different variant of the HERA feed in which the resonance is absent but whose primary beam properties are still being investigated.}.

Since the feed properties dominate the antenna's long-term response, we investigate the voltage response (Fig.~\ref{fig:SimulationResults}) and the power kernel (Fig.~\ref{fig:Kernels}) of the HERA antenna without the cylindrical skirt and a bare dipole. Removing the cylinder leads to a $\approx 10$\,dB improvement in the antenna response after $100$\,ns and eliminating the backplane as well yields a $20$ to $30$\,dB improvement. Hence, the delay response of the HERA antenna can potentially be improved by modifying the feed. There are several trade-offs in polarization and sensitivity that are made when we remove feed components which we now discuss.

\subsection{Tradeoffs in Beam Properties.}\label{ssec:Tradeoffs}
We determined above that the long-delay {\it knee} in the gain of the HERA dish is primarily caused by resonance in the cylindrical skirt and backplane that surround the PAPER dipole. These structures are added to  enhance the symmetry of its beam, to improve the effective area of the antenna, and reduce cross coupling\citep{DeBoer:2016}. It is therefore worth examining the impact that removing these components has on the properties of the HERA beam.

We first examine the trade-off that is made  in polarization when we remove elements of the original design. Faraday rotation is capable of imparting fine-scale frequency structure into the Q/U polarized Stokes visibilities.  Antennas in which the $X$ and $Y$ polarized beams are not identical can leak stokes Q/U visibilities and the spectral structure they contain into the Stokes-I visibilities from which delay power spectra are formed, potentially masking the 21\,cm signal \citep{Jelic:2010,Moore:2013,Moore:2015}. It is shown in \citet{Moore:2013} and \citet{Moore:2015} that this leakage is given approximately by 
\begin{equation}
P_{leak} \approx \frac{\mathcal{A}_{-}}{\mathcal{A}_{+}} P_P \equiv \xi P_P,
\end{equation}
where $P_P$ is the linearly polarized power spectrum, and 
\begin{equation}
\mathcal{A}_{\pm} = \int d \Omega |A_{xx}(\bl) \pm A_{yy}(\bl) |^2,
\end{equation}
where $A_{xx}$ is the $x$ polarized beam and $A_{yy}$ is the the $y$ polarized beam. In Table~\ref{tab:BeamProperties} we give $\xi$ at 150\,MHz, computed for the three configurations of the HERA feed discussed in \S~\ref{ssec:Knee}. We see that the feed with the cylinder has the smallest ellipticity with $\xi \approx 3 \times 10^{-3}$ over much of the band. Polarization measurements from \citet{Asad:2015} indicate a polarized power spectra of $P_P \approx 10^3$\,mk$^2$ within the EoR window. Thus, an antenna with $\xi \approx 3 \times 10^{-3}$ would produce leakage of several mk$^2$ which is roughly an order of magnitude below typical reionization peak amplitudes of several tens of mK$^2$. However, the feed variant in which the cylinder has been removed has an ellipticity that is roughly $\approx 4 \times$ larger and a bare dipole, $\approx 10 \times$ larger. These enhanced ellipticities would bring the level of the polarization leakage to or above the level predicted in \citet{Asad:2015}. On the other hand, it is possible the LST binning and averaging over many nights with different rotation measures can suppress polarization leakage by a factor of $\approx 10^{3}$ in the power spectrum \citep{Moore:2015}. 

The second impact that we consider, is the effective area of the antenna, given by \citep{Wilson:2009},
\begin{equation}
\Aeff =\frac{\lambda^2}{\int d \Omega A(\bl)}. 
\end{equation}
In Table~\ref{tab:BeamProperties}, we see that the effective area at $150$\,MHz is actually enhanced by the removal of the cylinder but is negatively impacted by the removal of the backplane which introduces a significant back-lobe, reducing the beam's directivity. As detailed in \citet{Neben:2016}, this reduction in the effective area of the dish lowers HERA's sensitivity by a factor of order unity.

\begin{table}
\caption{Ellipticity and effective area of the HERA antenna for different feed configurations.} 
\begin{tabular}{l|c|c} 
Feed Configuration & Ellipticity, $\xi$ & Effective Area, $\Aeff$, (m$^2$) \\ \hline
Cylindrical and Backplane & $3.5 \times 10^{-3}$ & 56.7  \\
No Cylinder & $1.5 \times 10^{-2}$ & 97.2  \\
No Cylinder or Backplane & $2.8 \times 10^{-2}$ & 41.9 
\end{tabular}
\label{tab:BeamProperties}
\end{table}

The development of the HERA feed is still underway and it is possible that improvements in the feed will allow for the 150\,MHz resonance to be eliminated while maintaining the illumination and beam-symmetry of the original feeds.


\subsection{Verifying Our Framework with $S_{11}$ Measurements and Simulations}\label{ssec:S11}

We now assess the accuracy of our time-domain simulation framework by comparing simulations to measurements of the $S_{11}$ parameter of the HERA dish. Up until now, our simulations have derived the voltage response of the dish using simulations of an incoming plane wave as is the case for radio signals arriving from objects at cosmological distances. It is possible to probe the gain of the dish using objects in the far field such as known radio sources \citep{Thyagarajan:2011,Pober:2012,Colgate:2015} or constellations of ORBCOMM satellites \citep{Neben:2015,Neben:2016}. However natural radio sources are too weak to probe the dish response at the $\lesssim10^{-4}$ level necessary to verify our simulations and the ORBCOMM technique can only be used to map the gain at the 137\,MHz ORBCOMM transmission frequency. Work is currently underway to use broad-band transmitters flown into the far field of the dish by drones \citep{Jacobs:2016} but this system is still under development. Reflectometry of the dish with a VNA is a straightforward alternative used in \citet{Patra:2016} to estimate the gains directly. Rather than comparing their estimate of the gain with our predicted gains (which is done in their paper), we set up a time domain simulation to compute the $S_{11}$ parameter of HERA antenna and compare it to direct $S_{11}$ measurements.
%
%
\begin{figure}
\includegraphics[width=.5\textwidth]{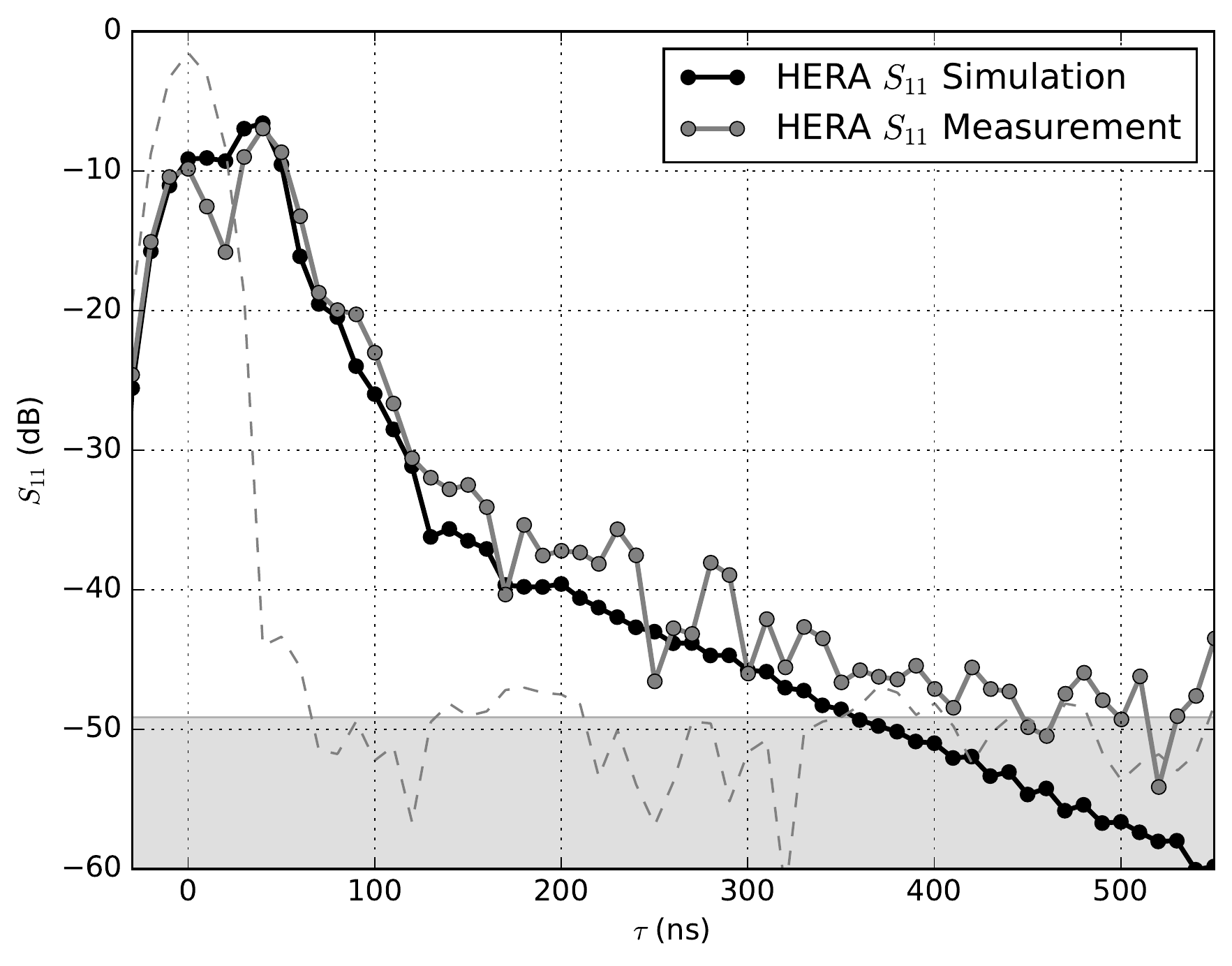}
\caption{A comparison between time-domain simulations (black line) and measurements (grey line) of $S_{11}$ for the HERA dish. We also show an $S_{11}$ measurement with the cables leading from the VNA to the feed terminated by an open circuit which allows us to probe the dynamic range of the measurement. We use the standard deviation of the open measurement (grey dashed line) between $200$ and $400$\,ns as our systematic floor (grey shaded region). We find very good agreement between our $S_{11}$ measurement and the simulation, validating the predictions of our simulations. Both the simulations and measurements in this figure were derived from delay transforms over 100\,MHz.}
\label{fig:S11}
\end{figure}
 
$S_{11}$ refers to the complex ratio between a voltage signal transmitted into the feed terminals and the voltage reflected back as a function of frequency, $ S_{11}(f) \equiv v_\text{trans}(f)/v_\text{recv}(f) $. Measurements, described in further detail in \citet{Patra:2016}, of $S_{11}$ were taken on a prototype HERA dish at the National Radio Astronomy Observatory Green Bank facility using an Agilent 8753D VNA. The VNA was connected to the antenna's balanced feed terminal via a 30.5 m length of RG-8X-LL coaxial cable and M/A Com HH-128 180 degree hybrid junction.  The VNA was calibrated at the balanced end of the hybrid junction using a set of termination standards having SMA connectors.  A small adapter from SMA to the post terminals on the feed was not included in the calibration.  The VNA excites the terminals with a band-limited (100-200\,MHz) pulse, and the complex reflected voltage at the calibration plane is measured as a function of frequency.  The reflection for the case of open terminals was measured to confirm the dynamic range and resolution of the measurement. While our simulations are terminated with an impedance of $125$\,$\Omega$, the VNA measurements are terminated at $100$\,$\Omega$ which only has a small effect on the large-delay response of the antenna (Fig.~\ref{fig:ImpedanceCompare}). 

Because $S_{11}$ is defined as a ratio in the frequency domain in a form identical to our voltage gains, we may run time-domain simulations similar to those described in \S~\ref{ssec:Simulations} except rather than simulating an incoming plane wave, we simulate the excitation of the feed terminals by a delta-gap impedance port and record output voltage as a function of time. We calculate $S_{11}$ from the simulation in the same manner that we obtained $\ftr$ using equation~\ref{eq:Inversion} with $\ftv_\text{recv}$ taking the place of $\ftv$ in the numerator inside the Fourier transform and $\ftv_\text{trans}$ taking the place of $\fts$ in the denominator. 

In Fig.~\ref{fig:S11} we show the simulated amplitude of $S_{11}$ as a function of frequency for the HERA dish, observing a distinctive two peaked structure before a steep die off in delay that transitions to a shallower falloff at $\approx 150$\,ns. The first peak is due to the reflection of the input wave off of the back of the feed while the second, roughly $35$\,ns later arises from the transmitted component of the input wave reflecting off of the dish and arriving back at the feed. The ensuing long term die off arises from reflections within the feed and dish structure and for reasons that will be elaborated on in \citet{Patra:2016} corresponds very closely to our simulations of the dish gain itself (compare with Fig.~\ref{fig:SimulationResults}).  We get a sense of the dynamic range of the measurement by unhooking the SMA adapter that attaches the sleeved dipole feed to the cable from our VNA, forming an open circuit that should ideally give a reflection coefficient of $\approx 1$ at zero delay with no reflections at any other times. In this measurement, we find noise-like structure at $\approx -50$\,dB. Below $\approx 500$\,ns, we found that this structure does not integrate down with time, leading us to conclude that it is caused by systematics, likely uncalibrated low level reflections in the VNA-feed cables. The level of this noise sets the systematic floor in our measurements which we show as a grey shaded region in Fig.~\ref{fig:S11}. We find that in the region where the $S_{11}$ measurement is above the systematics floor, there is good agreement with our simulations (within several dB).

\section{The Effect of the Spectral Structure due to Reflections and Resonances in the HERA dish on Foreground Leakage and Sensitivity}\label{sec:Sensitivity}
We can now explore the impact of the dish's performance on the leakage of foregrounds out of the wedge, and into the EoR window. Beyond the delay kernels considered in this paper and \citet{Patra:2016}, the extent of leakage will depend both on the angular structure of the primary beam, which is established through measurements and simulations in \citet{Neben:2016} and the brightness of the foregrounds themselves \citet{Thyagarajan:2016}. In this section, we investigate the amplitude of foreground leakage as a function of delay given the angular primary beam model and our simulation of the dish's spectral structure. We start by performing a physically motivated extrapolation of the delay structure observed in our simulations to cover the range of comoving scales relevant to power spectrum measurements (\S~\ref{ssec:Extrapolation}). In \S~\ref{ssec:Leakage}, we combine this extrapolated delay response with simulations of foregrounds to determine the overall level of foreground power in HERA's visibilities. This leakage will cause large-scale line-of-sight Fourier modes to be contaminated by foregrounds and hence inaccessible to the foreground filtering approach. Since the signal-to-noise ratio is maximized at the smallest $k$ values, the loss of these modes will reduce the significance of the power spectrum detection and negatively impact the overall bottom line of the science that HERA can accomplish. We explore the impact of the chromaticity due to reflections on science using the Fisher matrix formalism in \S~\ref{ssec:Science}.

\subsection{Extrapolating the Bandpass and Power Kernel}\label{ssec:Extrapolation}

Our simulations of the dish response only extend to $\approx 400$\,ns. However, interferometers are expected to be most sensitive to the 21\,cm power spectrum at comoving scales between $k\approx 0.1-0.5$ $h$\,Mpc$^{-1}$, corresponding to the delays between 180 and 900\,ns at $z=8$. To extrapolate out to the higher delays in this range, we assume that the response function is dominated by a sum of field oscillations within the feed's cylindrical skirt and reflections between the feed and the dish.
\begin{equation}
\widetilde{r}=\widetilde{r}_o + \widetilde{r}_r
\end{equation}
 The long term falloff after the knee, which is dominated by oscillations within the cylinder, appears as a line on a linear-log plot (Fig.~\ref{fig:SimulationResults}). This is the response we would expect for a damped harmonic oscillator. We thus model the voltage response from the cylindrical skirt as an exponential
\begin{equation}\label{eq:RefOscillator}
\widetilde{r}_o = A_o X_o^{-\frac{\tau 150\,\text{MHz}}{2Q} }
\end{equation}
where $Q$ is the quality factor of the cavity which we determine to be $\approx 6.5$. 

We see in Fig.~\ref{fig:SimulationResults} that when the cylinder and backplane are removed, the response of the antenna falls off much faster, in line with the response below $\approx 130$\,ns. Like the response from the cylinder, this falloff also goes exponentially which we now show is also indicative of reflections.

 We let $\Gamma_d$ represent the reflection coefficient of the dish vertex and $\Gamma_f$ represent the reflection coefficient of the feed in the presence of each other. An electromagnetic wave incident on the feed, at $t=0$, is accepted with amplitude $(1+\Gamma_f)$. The reflected component travels back to the dish and acquires an amplitude of $(\Gamma_f \Gamma_d)$ before returning at time, $\tau_d$ later where $(1+\Gamma_f)$ will be accepted and $\Gamma_f$ will be reflected back towards the dish. Summing the infinite series of reflections, the time dependent voltage at the feed is
\begin{equation}
\widetilde{v}_r(t) = (1+\Gamma_f)\sum_m \left( \Gamma_f \Gamma_d \right)^m \widetilde{s}(t-m \tau_d),
\end{equation}
which implies that
\begin{equation}\label{eq:deltaSeries}
\widetilde{r}_r(\tau) = (1+\Gamma_f) \sum_m \left( \Gamma_f \Gamma_d \right)^m \delta_D(\tau-m\tau_d).
\end{equation}
In practice, each delta-function in equation~\ref{eq:RefModel} is convolved with the window function arising from the finite bandwidth of the instrument. For our simulations, which have a bandwidth of $100$\,MHz, the width of the window function is $\approx 10$\,ns, partially filling in the gaps between each reflection. Since the number of reflections is $m=t/\tau_d$, than we can write the long-term delay response in discrete form as 
\begin{equation}
\widetilde{r}(n d \tau) \approx (\Gamma_f \Gamma_d)^{n d\tau/\tau_d}
 \end{equation}
which is an exponential in time. We thus model the response due to reflections between the feed and the dish as an exponential
\begin{equation}\label{eq:RefModel}
\widetilde{r}_r = A_r X_r^{(\tau/30\text{ns})}.
\end{equation}

To extrapolate beyond $400$\,ns, we fit the voltage response to the sum of two exponentials for $\widetilde{r}_o$ and $\widetilde{r}_r$ and set the feed response beyond the maximum time our simulation to this power law sum. In our pedagogical treatment, we have assumed that the reflection coefficients are frequency independent which is not the case in real life. The impact of frequency dependent reflection coefficients is to replace the summands in equation~\ref{eq:deltaSeries} with the Fourier transform of $(\Gamma_f\Gamma_d)^m$ evaluated at $t=(\tau - m \tau_d)$. As long as $(\Gamma_f\Gamma_d)^m$ is compact in delay space, which is the case if the reflection coefficients evolve smoothly with frequency, than our pedagogical approximation still yields the power law in equation~\ref{eq:RefModel}. 


\subsection{The Impact of delay response of the HERA Antenna on Foreground Contamination}\label{ssec:Leakage}

Given the chromaticity due to reflections and resonances in the HERA antenna, what Fourier modes will still be accessible with the delay filtering technique? To answer this question, we combine our extrapolated simulations of the HERA dish's spectral structure with simulations of foregrounds. 

The foreground model is discussed in detail in \citet{Thyagarajan:2015a} but we describe them briefly here for the reader's convenience. It consists of two major components: diffuse synchrotron emission from our Galaxy whose structure is described by the Global Sky Model (GSM) of \citet{deOliveiraCosta:2008} and a population of point sources including objects from the NRAO Sky Survey (NVSS) \citep{Condon:1998} at 1.4\,GHz and the Sydney University Mologolo Sky Survey (SUMMS) \citep{Bock:1999} at 843\,MHz. Their fluxes are extrapolated to the observed $100$-$200$\,MHz band using a spectral index of $\langle \alpha \rangle=-0.83$ determined in \citet{Mauch:2003}. Visibilities are computed from the diffuse and point source models assuming an achromatic angular response for the HERA beam at $137$\,MHz described in \citet{Neben:2016}. We compute two sets of visibilities: one in which the spectral structure of the dish is assumed to be completely flat, and another in which the beam is multiplied by the frequency dependent gain at zenith determined by our simulations.  

\begin{figure}
\includegraphics[width=.5\textwidth]{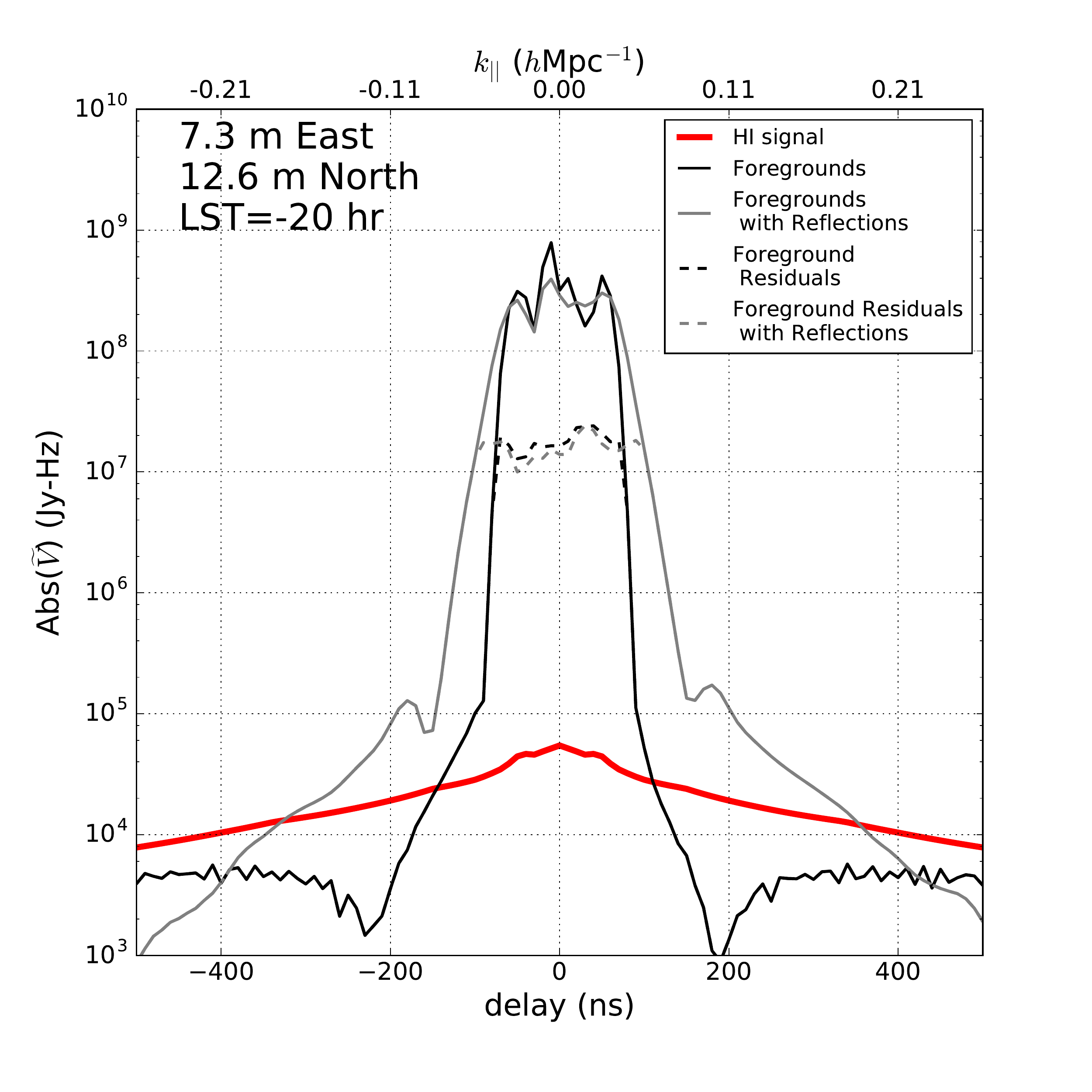}
\caption{The absolute magnitude of a  100-200\,MHz delay transformed visibility from a 14-meter baseline (blue line) compared to the same visibility (green line) contaminated by the delay-response observed in our simulations of the HERA dish. We see that the extended delay kernel smooths out structure, originating from foregrounds, within the horizon. For HERA, we expect to use the delay-CLEAN to remove foregrounds. However, the depth of CLEANing is limited by the noise level on a single baseline. We show the foreground residuals from a CLEAN down to the $5\,\sigma$ noise level after 20 minutes of integration. Since CLEANing cannot distinguish between foregrounds and signal, it should only be performed within a narrow region of delay-space, close to the horizon and cannot remove the broad wings leaked by the resonance unless it is accurately modeled.}
\label{fig:Cleaning}
\end{figure}

The foreground filtering procedure employed by PAPER and HERA involves delay transforming the visibilities and performing a 1D CLEAN \citep{Parsons:2009,Parsons:2012b} which discovers and subtracts foregrounds within the horizon plus a small buffer, allowing for the suppression of foreground side-lobes in delay space. The level of foreground subtraction possible by this procedure is limited by the thermal noise level on the visibility, which in turn depends on the number of time steps and redundant baselines that are averaged before performing the CLEANing step. In this work, we assume that each visibility is CLEANed independently with a twenty minute cadence. Each short baseline is coherent for $\approx 10$\,s per night \citep{Ali:2015} so this is equivalent to each local sidereal time (LST) being integrated for $\approx 120$ nights before CLEANing. If we were to average over redundant baselines this number would drop enormously. The standard deviation on the real and imaginary part of a single delay transformed visibility is \citep{Morales:2004}
\begin{equation}
\sigma_V = \frac{\sqrt{2 B} k_B T_{sys}}{A_e \sqrt{\tau}},
\end{equation}
where $A_e$ is the effective area of the dish, $B$ is the bandwidth, $T_{sys}$ is the system temperature, $\tau$ is the integration time, and $k_B$ is the Boltzmann constant. For $T_{sys}$ we use the equation $T_{sys} = 100\text{K} + T_{sky}$ where $100$\,K is the temperature of the PAPER receiver and $T_{sky} = 60 (\lambda/1\,\text{meter} )^{2.55}$ is the sky temperature \citep{Rogers:2008,Fixsen:2011}. For $A_e$ we use the value of $98$\,m$^2$ determined in \citet{Neben:2016}. We CLEAN down to five times the thermal noise level. Subtracting a model of the sky is an alternative to CLEANing that could reach below the noise level on a single visibility but also assumes a priori knowledge of the foregrounds and the instrumental response. In Fig~\ref{fig:Cleaning} we compare the delay transform of visibilities before and after CLEANing at the LST of 4 hours. While CLEANing is able to remove structure within the horizon, it does not reduce any of the power leaked outside of the horizon by the chromaticity of the dish. If we knew the spectral response of the dish perfectly, then we might be able to CLEAN with this kernel and remove the supra-horizon structure though the true direction-dependent nature of the chromaticity would complicate this task.

To form estimates of the 21\,cm power spectrum, we split each visibility into Blackman-Harris windowed sub-bands centered at redshift intervals of $\Delta z = 0.5$ and each with a noise equivalent bandwidth of $10$\,MHz\footnote{The end to end width of each Fourier transformed interval is 20\,MHz, however the noise equivalent width of this interval (given by the integral of the bandpass squared) is only $\approx 0.5$ the full width since the Blackman-Harris window suppresses the edge channels.}. The flat sky approximation allows us to Fourier transform each interval in frequency, square, and multiply by a set of prefactors to obtain a power spectrum estimate \citep{Parsons:2012a,Parsons:2014},
\begin{equation}\label{eq:PS}
\widehat{P}({\bf k}) = \left( \frac{2 k_B}{\lambda^2} \right)^2 \frac{X^2 Y}{B_{pp} \Omega_{pp}} | \widetilde{V}({\bf u}) |^2.
\end{equation}
Here, $\lambda$ is the central wavelength of the observation, $k_B$ is the Boltzmann constant, $B_{pp}$ is the integral of the square of the bandpass, and $\Omega_{pp}$ is the integral of the primary beam squared over solid angle. $X$ and $Y$ are linear factors converting between native interferometry and cosmological coordinates, defined through the relation $2 \pi {\bf u}=2 \pi (u,v,\eta) = (X k_x, X k_y, Y k_z)$. 

As a drift scan instrument, HERA will observe the sky at many LSTs within the declination stripe that passes through its primary beam, averaging over the power spectrum estimate at each LST. It is well documented that foreground power varies significantly over LST \citep{Thyagarajan:2015a}, hence such an estimate will either filter or weight LSTs in a way that minimizes the impact of the most contaminated observations. For our analysis, we focus on a single LST of 4 hours which is a patch of sky with low Galactic contamination. Such a patch is representative of the kind that HERA will focus most of its observing time on.

\begin{figure*}
\includegraphics[width=\textwidth]{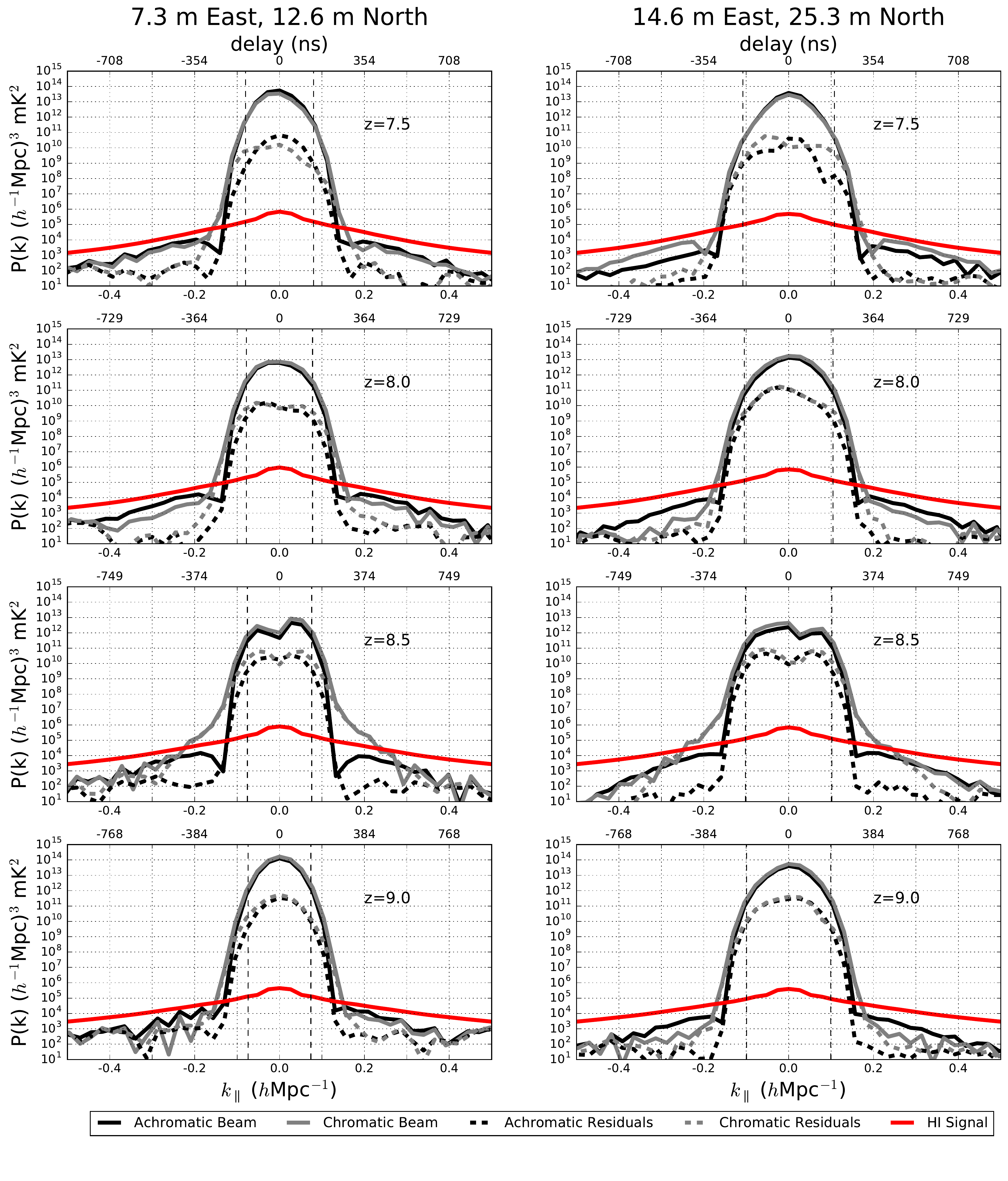}
\caption{The delay transform power spectrum for different baselines over several redshifts with (green solid line) and without (blue solid line) the presence of the simulated chromaticity in the HERA dish. Each estimate is computed using the square of a Blackman-Harris windowed  delay transform (equation~\ref{eq:PS}) over a noise equivalent bandwidth of $10$\,MHz. We also show the power spectra of foreground residuals after CLEANing the entire $100$\,MHz band with (dashed cyan line) and without (red dashed line) the simulated spectral structure. For all subbands, except $z=8.5$, we find that the delay response of the HERA antenna has a negligible effect on the $k_\parallel$ where the foregrounds drop below the signal level. This is the same subband where enhanced spectral structure due to resonance in the feed cylinder is observed in our simulations (Figs.~\ref{fig:KernelsSubbands} and \ref{fig:Residuals}). Outside of the resonance, the chromaticity of the antenna, which is primarily sourced by feed-dish reflections, is practically identical to the achromatic model. It follows that feed-dish reflections have a negligible impact on HERA's 21\,cm power spectrum sensitivity. For the achromatic beam, power outside of the horizon is leaked by the window-function of the finite-band Fourier transform. Thus, reducing power inside of the horizon does reduce the level of contamination outside of the window except when the structure is intrinsic to the frequency gain of the antenna as is the case at z=8.5 where structure is present at super-horizon delays due to the feed resonance.}
\label{fig:BothBaselines}
\end{figure*}

Computing the power spectra, we inspect the amplitude of foregrounds given the delay-response and angular pattern of the HERA dish for baselines of two different lengths in Fig.~\ref{fig:BothBaselines}. In both baselines, we find that the residuals after CLEANing tend to be at similar levels except at the subband centered at $z=8.5$ (150\,MHz) where the cylinder resonance is present. In bands outside of the resonance, foreground power outside of the horizon is dominated by side-lobes of the finite window function used in the Fourier transform. Hence, if we remove power inside of the horizon through cleaning, the level of this leakage is reduced. Power that is not reduced by the 1d clean is caused by intrinsic super-horizon structure in the antenna gain. Over the resonance, foreground residuals remain above the signal level out to $k_\parallel=0.23$\,$h$Mpc$^{-1}$. Several other baselines especially those oriented entirely in the E-W direction, which we do not show, contain up to two orders of magnitude greater galactic  contamination as is noted in \citet{Thyagarajan:2015b}. In data analysis, these baselines would be down-weighted or discarded so that they do not bias our final estimate. Outside of the resonance, beam chromaticity is dominated by beam-antenna reflections and since the level of foreground contamination is nearly identical to the achromatic beam, we conclude that feed-dish reflections do not pose a significant limitation HERA's ability to measure the EoR power spectrum. Dishes are therefore a viable strategy for scaling the collecting area of 21\,cm interferometry experiments at low cost. 

The level of the foregrounds in Fig.~\ref{fig:BothBaselines} is conservative since no attempt has been made to apply inverse covariance weighting techniques \citep{Tegmark:1997a,Liu:2011,Dillon:2013,Parsons:2014,Liu:2014a,Liu:2014b,Dillon:2015a,Dillon:2015b,Trott:2016} or fringe rate filtering \citep{Parsons:2015}. Applications of inverse covariance filters in recent PAPER observations have yielded reductions in foreground power by greater than an order of magnitude \citep{Ali:2015}, hence our simulations show that even with the presence of the enhanced spectral structure from reflections, HERA will be able to isolate foregrounds well below the level of thermal noise in most of the EoR window. 

\subsection{The Implications of Reflections and Resonances on EoR Science}\label{ssec:Science}
A primary near-term goal of 21\,cm EoR observations is to obtain information about the nature of the sources that drove reionization. Since the amplitude of the 21\,cm signal is largest at smaller $k$ values, a loss of large scale signal due to foreground contamination eliminates the modes on which HERA would otherwise have the greatest signal to noise detections, reducing its overall sensitivity. In this section, we use the Fisher Matrix formalism to estimate the impact of the HERA's intra-dish reflections on its sensitivity and ability to constrain the astrophysics of reionization. The Fisher Matrix allows us to forecast the covariances and errors on reionization parameters given errors on power spectrum observations due to the uncertainties caused by cosmic variance and thermal noise which is in turn determined by the $uv$ coverage and observing time of the interferometer. The covariance between the parameters, ${\bf \theta}$, of an astrophysical model is given by the inverse of the Fisher matrix, ${\bf F}$ which for Gaussian and independently determined power spectrum bins may be written approximately as \citep{Pober:2014},
\begin{equation}\label{eq:Fisher}
F_{ij} \approx \sum_{k,z} \frac{1}{\sigma^2(k,z)} \frac{\partial \Delta^2(k,z)}{\partial \theta_i} \frac{\partial \Delta^2(k,z)}{\partial \theta_j},
\end{equation}
where $\Delta^2(k,z)$ is the power spectrum amplitude for some $k-z$ bin and $\sigma^2(k,z)$ is the variance of the power spectrum estimate. We write this equation as approximate since it ignores additive terms arising from the dependence of $\sigma^2(k,z)$ on model parameters due to cosmic variance which only contribute at the $\approx 1$\% level \citep{EwallWice:2015b}. 

To simulate $\Delta^2$, we use the publicly available {\tt 21cmFAST}\footnote{\url{http://homepage.sns.it/mesinger/DexM___21cmFAST.html}} code \citep{Mesinger:2011} which generates realizations of the 21\,cm brightness temperature field using the excursion set formalism of \citet{Furlanetto:2004}. We employ a popular three parameter model of reionization \citep{Mesinger:2012} with the following variables 
\begin{itemize}
\item ${\bf \zeta}$: The ``ionization efficiency" is defined in \citet{Furlanetto:2004} to be the inverse of the mass collapse fraction necessary to ionize a region and is computed from a number of other physical parameters including the fraction of collapsed baryons that form stars and the UV photon escape fraction. Because $\zeta$ acts as an efficiency parameter, its primary effect is to change the timing of reionization. We choose a fiducial value of $\zeta=20$, though expected values range anywhere between $5$ and $50$ \citep{Songaila:2010}. 
\item ${\bf R_\text{mfp}}$: The presence of Lyman limit systems and other potential absorbers within HII regions causes UV photons to have a finite mean free path denoted by $R_\text{mfp}$. In the {\tt 21cmFAST} framework, HII regions cease to grow after reaching the radius of $R_\text{mfp}$, primarily impacting the morphology of the signal. We choose a fiducial value of $R_\text{mfp}=15$\,Mpc which is in line with recent simulations accounting for the subgrid physics of absorption \citep{Sobacchi:2014}. 
\item ${\bf T_\text{vir}^\text{min}}$: The minimal mass of dark matter halos that hosted ionizing sources. While in principle, halos with virial temperatures as small as $10^2$\,K are thought to be able to form stars \citep{Haiman:1996a,Tegmark:1997c}, thermal and mechanical feedback have been seen to raise this limit to as high as $10^5$\,K \citep{Springel:2003,Mesinger:2008,Okamoto:2008}. We choose a fiducial value of $T_\text{vir}^\text{min} = 1.5\times 10^4$\,K which is set by the atomic line cooling threshold. 
\end{itemize}
In order to account for the degeneracies in the power spectrum between heating from X-rays and reionization from UV photons, we also marginalize over three additional parameters that describe the impact of heating from early X-ray sources as explored in \citep{EwallWice:2015b}. These are the X-ray heating efficiency, $f_X$; the maximal energy of X-ray photons that are self absorbed by the ISM of early galaxies, $\nu_\text{min}$; and the spectral slope, $\alpha$ which are taken to have fiducial values of $1$, $0.3$\,keV, and $-1.2$ respectively. We choose to parameterize our model in terms of the fractional difference of each variable from its fiducial value so that, for example, $\theta_\zeta = (\zeta - \zeta_\text{fid})/\zeta_\text{fid}$ and compute the derivatives in equation~\ref{eq:Fisher} by performing a linear fit to realizations of the 21\,cm power spectrum calculated by {\tt 21cmFAST} at $\theta_i= \pm 10^{-2}, \pm 5\times 10^{-2}, \pm 10^{-1}$, and $ \pm 2 \times 10^{-2}$. 

For each measurement in the $uv$ plane, the standard deviation of a power spectrum measurement ($\sigma^2(k,z)$) is given by the direct sum of sample variance and thermal noise \citep{McQuinn:2006} which in turn depends on the primary beam of the instrument and the time spent sampling each $uv$ cell. For our analysis, we assume that the $uv$ plane is sampled by circular apertures with effective areas of $98$\,m$^2$ and that $\tau({\bf k})$ is determined by a drift scan in which non-instantaneously redundant baselines are combined within each $uv$ cell. We compute the standard deviation of each $(k,z)$ bin for the proposed 350-element deployment of HERA (HERA-350) using the public {\tt 21cmSense}\footnote{\url{https://github.com/jpober/21cmSense}} code \citep{Pober:2013b,Pober:2014}.

\begin{figure}
\includegraphics[width=.5\textwidth]{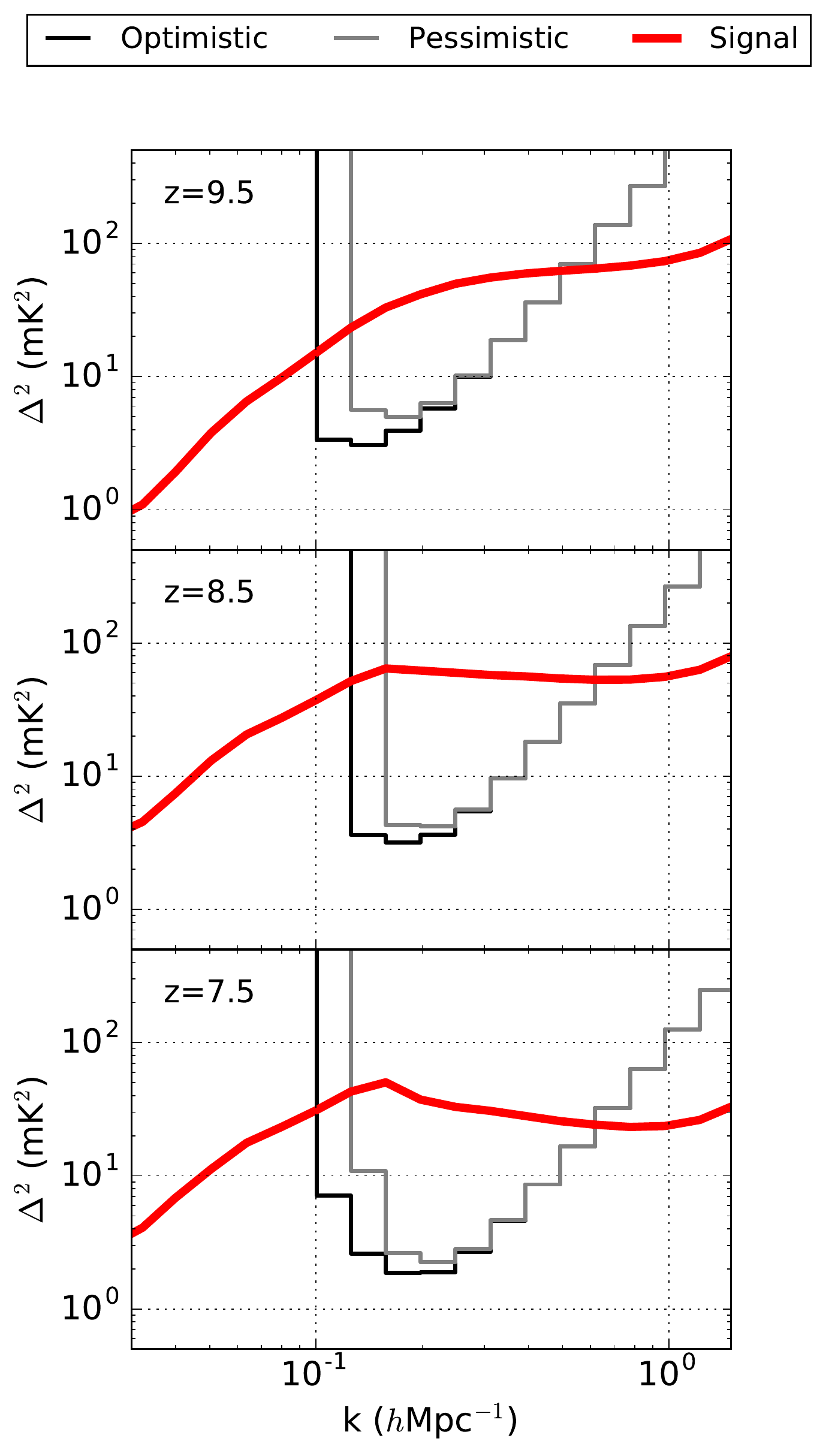}
\caption{The $1\sigma$ thermal noise levels achieved by HERA-350 at three different redshifts with (grey line) and without (black line) the presence of beam chromaticity due to the chromaticity observed in this work. We compare these noise levels to the fiducial power spectrum signal (red line). We saw in Fig.~\ref{fig:BothBaselines} that with reflections, foregrounds exceed the signal level out to $k=0.23$\,$h$Mpc$^{-1}$ at $z=8.5$ which we assume are unusable, forcing us to ignore modes out to $350$\,ns beyond the horizon, leading to the sensitivity projected in the red curve. The absence of these reflections allows us to work within $250$\,ns of the horizon (green curve), leading to an increase in sensitivity by a factor of $\approx 1.5$. }
\label{fig:Sensitivity}
\end{figure}

\begin{figure}[h!]
\includegraphics[width=.5\textwidth]{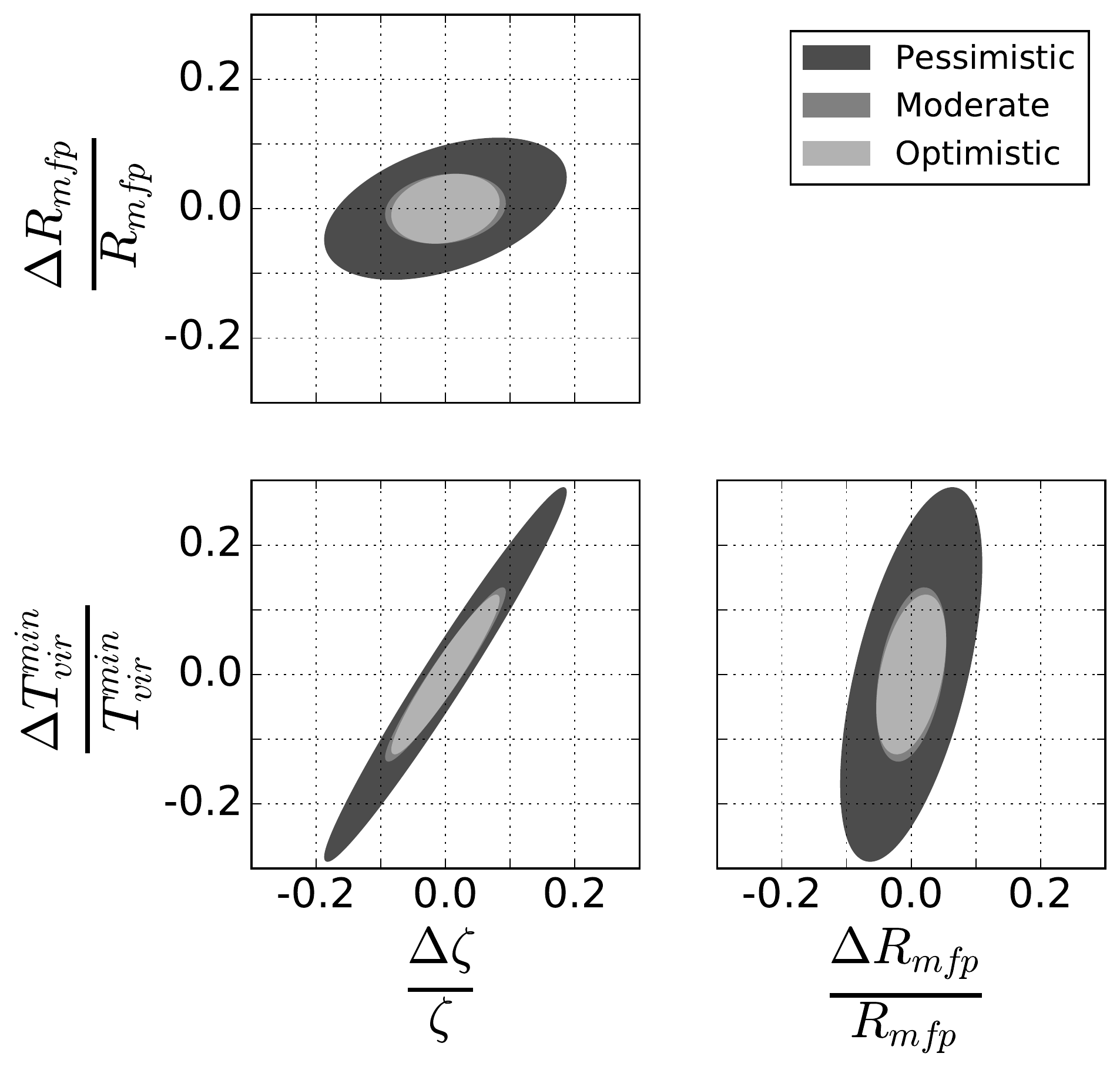}
\caption{95\% confidence regions for reionization parameters assuming $1000$ hours of observation on HERA-350 between the redshifts between 7.5 and 12. The presence of the reflections leads to an increase in the major axes of these confidence regions by a factor of one to two.}
\label{fig:Confidence}
\end{figure}

\begin{figure*}
\includegraphics[width=\textwidth]{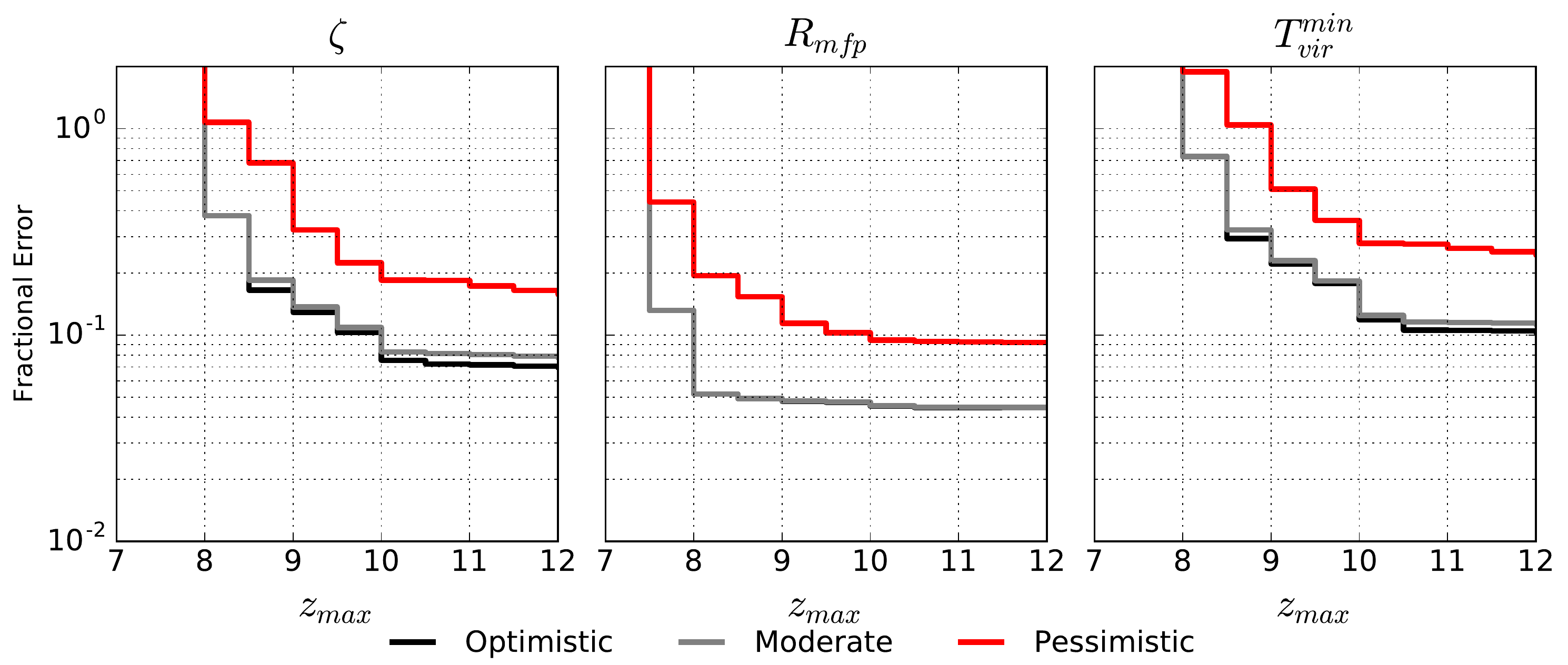}
\caption{Fractional Errors on reionization and heating parameters as a function of maximal observed redshift out to the low end of HERA-350's initial observing band at $z=12$. The presence of strong reflections contained within a small subband at $z=8.5$ has a minimal impact on our overall constraints on reionization parameters. If these reflections are not localized they can worsen our sensitivity to some parameters such as $T_\text{vir}^\text{min}$ by a factor of two.}
\label{fig:Errors}
\end{figure*}

We have seen in Fig.~\ref{fig:BothBaselines} that the simulated chromaticity of the dish leaks foregrounds beyond the horizon to varying degrees depending on the subband with the resonance occuring at $z=8.5$(150\,MHz). In this subband, foregrounds exceed the signal out to $\sim 380$\,ns which for a $14.6$\,m baseline is $\approx 330$\,ns beyond the horizon. At all other redshifts, the leakage due to reflections only extends to $\approx 250$\,ns beyond the horizon. To determine the impact of the observed reflections on HERA's ability to constrain the astrophysics of reionization, we vary the cutoff in delay below which cosmological Fourier modes will be unobservable. To determine this cutoff, we consider three different scenarios for beam chromaticity that capture a range of possibilities informed by our simulations.

\begin{itemize}
\item {\bf Optimistic: The cylinder resonance is mitigated.} In our most optimistic scenario, we assume that a more advanced feed design is able to mitigate the 150\,MHz resonance in the cylinder while preserving polarization symmetry and effective area. In this case the foregrounds pass below the level of the signal at $250$\,ns beyond the horizon at all redshifts between $z=7$ and $12$, consistent with what is observed in the bands where the reflections are less severe or when they are not present at all. Since the achromatic beam stays above the foregrounds below $\approx 250$\,ns as well, this number is set by the width of the Blackman-Harris window function and more optimal estimators of the power spectrum may reduce it substantially (see \citet{Ali:2015}). Hence, even our optimistic scenario is on the conservative side, making it more pessimistic than previous Fisher analyses \citep{Pober:2014,Liu:2015a,Liu:2015b,EwallWice:2015b}. In this scenario, we assume that only modes within 250\,ns of the horizon are unobservable. 
 
\item {\bf Moderate: The Simulations accurately capture the chromaticity of the dish.} In this case, we assume that the impedance match is sufficient for the contamination from reflections to pass below the level of the signal at $250$\,ns beyond the horizon except at redshift $8.5$ where the feed resonance causes foregrounds to extend above the signal to $350$\,ns beyond the horizon. In this scenario, we assume that modes within 250\,ns of the horizon are unobservable except for redshift 8.5 where modes below 350\,ns beyond the horizon are unobservable. This is the scenario that would follow directly from our simulation results.

\item {\bf Pessimistic: Structure is present in all sub-bands.} In this scenario, we assume that the resonance is still present in the neighborhood of $150$\,MHz and that a very poor impedance match (similar to the 500\,$\Omega$ case in Fig.~\ref{fig:ImpedanceCompare}) is obtained, allowing for significant contamination from reflections over the entire band, beyond the resonance. For this case, structure would be present out to $\approx 350$\,ns over the entire $100$-$200$\,MHz frequency range covered by HERA. In this scenario, we assume that modes within 350\,ns of the horizon are unobservable across all redshift intervals.  
 
\end{itemize}
In Fig.~\ref{fig:Sensitivity}, we compare the level of $1\sigma$ thermal noise for our optimistic and pessimistic scenarios to the amplitude of the 21\,cm signal at several different redshifts. The smallest $k$ modes lost to foreground contamination are the highest signal to noise measurements that HERA is expected to obtain. Their absence impacts sensitivity in two ways. By leading to a reduction in the maximal signal to noise ratio of $\approx 1.5$ and a reduction in the total number of modes that the instrument is able to measure. 

Folding our calculations of thermal noise and the derivatives of $\Delta^2$ into equation~\ref{eq:Fisher} and inverting, we obtain the covariance matrix for model parameters. We show the 95\% confidence ellipses for the reionization parameters in Fig.~\ref{fig:Confidence}. The presence of the resonance within a limited sub-band about $z=8.5$ leads to an almost neglible increase in the extent of our confidence intervals while its presence across the entire band causes the lengths and widths of our confidence ellipses to increase by a factor of $\approx 2$. The diagonal elements of our covariance give us error bars on each parameter which we plot in Fig.~\ref{fig:Errors}. We see that similar to our confidence regions, the error bars on reionization parameters for our optimistic and moderate scenarios are nearly indestinguishable. In our worst case where the resonance is present across the entire band, we see an increase in our error bars by a factor of $\approx 2$. 

The error bars on reionization parameters, even for our most optimistic model, are a factor of a few larger than the ``moderate" errors predicted in previous works using likelihood analyses such as \citet{Pober:2014}, \citet{Greig:2015,Greig:2015b,Greig:2015c}  \citet{EwallWice:2015b}, \citet{Liu:2015a,Liu:2015b}. There are several reasons for this. Firstly, in our most optimistic scenario, we assumed that foregrounds cause the signal to be inaccessible below $250$\,ns beyond the edge of the wedge which corresponds to $k_\text{min} \approx 0.15$\,$h$Mpc$^{-1}$ at $z=8.5$ while in previous works a minimal comoving $k_\text{min} \approx 0.1$\,$h$Mpc$^{-1}$ was used. Secondly, previous studies assumed a fully illuminated HERA aperture, which for a $14$\,m diameter dish predicts an effective area of $\approx155$\,m$^2$. Electromagnetic simulations and ORBCOMM mapping of the angular beam pattern of the HERA dish show that the effective area of the antenna element is actually $\approx 98$\,m$^2$ at $137$\,MHz which leads to an increase in the overall thermal noise levels by a factor of $\approx 1.5$ \citep{Neben:2016}. 

Although conservative, our analysis shows that the level of the HERA antenna's delay response is not an insurmountable obstacle for the 21\,cm power spectrum measurement. Even if the additional systematics introduced by the feed resonance are present over the entire band, HERA will obtain a $\gtrsim 10 \sigma$ detection of the power spectrum and be capable of establishing precision constraints on the properties of the sources that reionized the IGM. 

\section{Conclusions}\label{sec:Conclusion}
In this paper, we have derived the impact of spectral structure from reflections in the analog signal chain of an interferometer on foreground contamination of the 21\,cm signal. We have used simulations of electromagnetic waves incident from zenith on the primary antenna element on HERA to determine the extent to which spectral structure due to reflections in the dish leak foregrounds into the EoR window. The results of our simulations of the dish's voltage response are broadly consistent with reflectometry measurements \citep{Patra:2016} and can be summarized in the following points. 
\begin{itemize}
\item Additional spectral structure, in excess of that observed in the antenna design of HERA's predecessor, PAPER, is observed in our simulations. Because this structure is present in simulations without the dish and is dramatically reduced when the feed's cylindrical enclosure is removed, we conclude that this structure arises primarily from a resonance within the feed cylinder. On the other hand, contamination from reflections between the feed and the parabolic dish, which are present outside of the 150\,MHz resonance, are relatively minor. Such reflections were perceived as a significant risk to the use of dish antennas in EoR studies and our simulations show that their impact on array sensitivity is small and 21\,cm experiments can be economically scaled up through the use of dishes. Further adjustments to the feed design aim to remove the cylindrical resonance while preserving collecting area and polarization match.

\item  The resonance which is responsible for the majority of contamination has a finite width of $\approx 10$\,MHz of the HERA band. Because estimates of the power spectrum are obtained from sub-intervals of $\approx10$\,MHz, the structure that we have simulated will only impact a single $\Delta z=0.5$ redshift interval.

\item Combining simulations of the HERA antenna's delay-response with the foreground model of \citet{Thyagarajan:2016}, we find that a resonance in the feed extends foregrounds above the level of the cosmological signal to $\approx 350$\,ns beyond the horizon while without the resonance (but with reflections), foregrounds extend above the signal to $\approx 250$\,ns beyond the horizon (Fig.~\ref{fig:BothBaselines}). These forecasts are conservative in that we do not attempt to inversely weight the foregrounds by their covariances or apply additional mitigation algorithms such as delay rate filtering. 

\item If beam chromaticity is contained within a $10$\,MHz subband around $150$\,MHz, then the overall constraints that HERA will be able to place on the astrophysics of reionization are minimally impacted. If the resonance were to extend across the entire band, our constraints on reionization parameters will suffer a two-fold increase in uncertainty but still remain on the order of $10$\%.
\end{itemize}

	
Our results are very encouraging since they show that the HERA design is capable of isolating the foregrounds to sufficiently low delays to make a high significance detection of the 21\,cm power spectrum during the EoR. Since reflections are not a major source of spectral structure, feeds suspended over dishes are a viable strategy for increasing the collecting area of HI interferometers at a modest cost.  Our results also emphasize the importance of mitigating any spectral structure in the analog signal chain. The contamination from the feed resonance masking our signal over some redshifts is mathematically identical to Fourier modes introduced by any instrumental spectral structure. Hence it is important that reflections be stringently suppressed at all steps in the signal chain beyond the dish to ensure that bright foregrounds fall below the EoR signal. A systematic definition of specifications on the spectral structure necessary to observe the EoR power spectrum at different $k$ values is presented in \citet{Thyagarajan:2016}. 

\section*{acknoledgments}
We would like to thank Danny Jacobs, Jonathan Pober, Gianni Bernardi, Peter Williams, and Jeff Zheng for their helpful comments. This work was supported by NSF grants AST-1440343 and AST-1410484,
the Marble Astrophysics Fund, and the MIT School of
Science. ARP acknowledges support from NSF CAREER
award 1352519. JSD acknowledges support from a Berkeley Center for Cosmological Physics Fellowship. AEW acknowledges support from
an NSF Graduate Research Fellowship under Grant No.
1122374.

\bibliographystyle{apj}
\bibliography{DishSimulation_paper}
\end{document}